\documentclass[10pt,journal,doublecolumn]{IEEEtran}
\usepackage{cite}
\usepackage{amsmath,amssymb,amsfonts}
\usepackage{algorithmic}
\usepackage{graphicx}
\usepackage{textcomp}
\usepackage{graphicx}
\usepackage[utf8]{inputenc}
\usepackage[margin=0.75in]{geometry}
\usepackage{graphicx}
\usepackage{cite}
\usepackage{amsmath}
\usepackage{cite}
\usepackage{amssymb}
\usepackage{wasysym}
\usepackage{comment}
\DeclareUnicodeCharacter{2061}{}

\DeclareUnicodeCharacter{0301}{}
\usepackage{enumerate}
\DeclareMathOperator*{\argmax}{argmax}
\usepackage{subfigure}
\usepackage{comment}
\usepackage[skip=10pt,font=scriptsize]{caption}

\usepackage{amsfonts}
\usepackage{authblk}
\usepackage{enumitem}
\usepackage{textcomp}
\def\BibTeX{{\rm B\kern-.05em{\sc i\kern-.025em b}\kern-.08em
    T\kern-.1667em\lower.7ex\hbox{E}\kern-.125emX}}
    
\begin{document}

\title{A Tutorial to Sparse Code Multiple Access}

\author{Saumya Chaturvedi,~Zilong Liu,~Vivek Ashok Bohara,~Anand Srivastava,~Pei Xiao
\thanks{Saumya Chaturvedi, Vivek Ashok Bohara, and Anand Srivastava are with Indraprastha Institute of Information Technology
(IIIT-Delhi), Delhi, New Delhi, 110020, India (e-mail: \{saumyac,
vivek.b, anand\}@iiitd.ac.in). Zilong Liu is with the School of Computer Science and Electrical
Engineering, University of Essex, Colchester CO4 3SQ, U.K. (e-mail:
zilong.liu@essex.ac.uk). Pei Xiao is with the Institute of Communication Systems, 5G Innovation Centre, University of Surrey, UK (E-mail: {p.xiao@surrey.ac.uk}).}}

\maketitle

\begin{abstract}
     Sparse Code Multiple Access (SCMA) is an enabling code-domain non-orthogonal multiple access (NOMA) scheme for massive connectivity and ultra low-latency in future machine-type communication networks. As an evolved variant of code division multiple access (CDMA), multiple users in SCMA are separated by assigning distinctive codebooks which display certain sparsity. At an SCMA receiver, efficient multiuser detection is carried out by employing the message passing algorithm (MPA) which exploits the sparsity of codebooks to achieve error rate performance approaching to that of the maximum likelihood receiver. Despite numerous research efforts on SCMA in recent years, a comprehensive and in-depth tutorial to SCMA is missing, to the best of our knowledge. To fill this gap and to stimulate more forthcoming research, we introduce the principles of SCMA encoding, codebook design, and MPA based decoding in a self-contained manner for layman researchers and engineers. 
\end{abstract}

\begin{IEEEkeywords}
Factor Graphs, Message Passing Algorithm, Sparse Code Multiple Access, Non-Orthogonal Multiple Access, Codebook Design.
\end{IEEEkeywords}

\section{Introduction}

%\subsection{Background}
The next-generation cellular systems are expected to support a wide range of vertical industries such as e-health, smart homes/cities, connected autonomous vehicles, and factories-of-future \cite{uses}. While this leads to ubiquitous digital data services anywhere and anytime, the explosive growth of communication devices and applications brings an increasingly congested spectrum which is one of the major challenges in the design of 5G communication networks and beyond. 

Multiple access, as one of the core techniques of wireless communication, aims to enable multiple users to access the finite resources simultaneously in an effective manner. Legacy multiuser communication systems mostly use orthogonal multiple access (OMA) schemes in which multiple users are orthogonal to each other with respect to certain type of resources. In the past  cellular networks, there have been time division multiple access (TDMA), frequency division multiple access (FDMA), code division multiple access (CDMA), and orthogonal frequency-division multiple access (OFDMA). As an example, in TDMA, each user is given a distinct time slot and no two users simultaneously share  the same time slot. In every OMA scheme, the number of users being simultaneously served is upper bounded by that of the orthogonal resources.

In recent years, non-orthogonal multiple access (NOMA) has attracted tremendous research attention with a major aim of providing massive connectivity, lower communication latency,  and higher spectral efficiency compared to OMA. %\cite{ad8bbbe54d5f43d88e864886dd63bd15}. 
In a NOMA system, two or more users are superimposed over an identical  physical resource (e.g., power, frequency, time, or code) to provide overloading factor larger than one \cite{Liu2021}.  

In 2013, sparse
code multiple access (SCMA) was proposed by H. Nikopour  and  H. Baligh as a disruptive code-domain NOMA scheme \cite{proposed}.  
SCMA is a generalized multiple access scheme building upon low-density signature CDMA (LDS-CDMA) \cite{lds}, where the latter is a special case of CDMA in which several bits of each user are mapped to a symbol followed by sparse sequence spreading. By contrast, QAM mapping and spreading are merged together in SCMA and therefore several incoming bits (of certain user) can be directly mapped to a sparse complex vector (codeword) which is drawn from its associated sparse codebook. %SCMA outperforms  LDS-CDMA in terms of its robustness against noise and multiuser interference due to the so-called constellation shaping gain introduced by carefully designed sparse codebooks. 

%\underline{\emph{Related Works:}}
\subsection{Related Works}
In general, there are two major research lines associated to SCMA: codebook (CB) design and multiuser detection (MUD). The goal of CB design is to design and optimize CBs with respect to certain channels (e.g., downlink vs uplink, Gaussian vs Rayleigh) in order to improve a number of system performance measures such as bit/packet error rate, spectral efficiency, and peak-to-average power ratio (PAPR). After the first CB design work \cite{cb2} in 2014, numerous research attempts have been made. In \cite{cb3}, new CBs are designed from \emph{M}-order pulse-amplitude modulation (PAM) to maximize the sum rate of SCMA. In \cite{interleaving}, the authors proposed a multi-dimensional CB based on constellation rotation and interleaving to maximize the minimum Euclidean distance between the codewords. In \cite{starqam}, novel CBs for both Rayleigh and Gaussian channels based on Star-QAM constellations have been developed with the aim of minimizing the error probability. Inspired by recent major advances in machine learning,  new CBs have been found in \cite{ref4} using deep-learning-based method by considering minimum bit error rate (BER) as the optimization criteria.  In \cite{gam}, %[17]
golden angle modulation
(GAM) constellations have been used for CBs which exhibit excellent error rate performances in both uplink and downlink Rayleigh fading channels. Recently, a low complexity construction algorithm for near-optimal CBs has been developed in \cite{nearopt}.
 
Thanks to the sparsity of the codebooks, the multiuser signals can be efficiently detected and recovered with the aid of  message passing algorithm (MPA) \cite{lds} whose error rate performance approaches to that of  maximum a posteriori (MAP) detector. In MPA, the belief messages are passed along the edges of the corresponding factor graph which is associated with the sparse codebooks of an SCMA system. It is shown in \cite{yiqun} that MPA works well even in an SCMA system with an overloading factor of three. In \cite{ref7}, a simpler MPA in log-domain is introduced by simplifying the exponent and multiplication operations to maximization and addition operations, respectively. In \cite{ref8}, a partial marginalization based MPA is proposed for fixed low-complexity SCMA detection. \cite{thresholdmpa} proposed a belief threshold-based MPA in which the belief messages updates for every user are checked in every iteration. This reduces the total number of computational operations when the number of iterations increases. In \cite{lutmpa}, a lookup table method is designed to simplify the function node update which ensures the stable convergence of MPA. In \cite{mimompa}, a hybrid belief propagation and
expectation propagation  receiver was proposed for a downlink MIMO-SCMA system. In \cite{edgempa}, a threshold-based edge selection and Gaussian approximation (ESGA) was proposed in which partial Gaussian approximation was applied dynamically for significant complexity reduction at the receiver.
%In \cite{ref7}, max log MPA was proposed which converts multiplication and exponent operations to addition and maximization operation using Jacobian logarithm formula. 
%\underline{\emph{Organization of this paper:}}
\subsection{Motivations and Contributions}
%Since SCMA was proposed in \cite{proposed}, number of works have been done  to improve the SCMA technique and make it more simpler for practical implementations. 
Albeit numerous research attempts on SCMA have been made in the past few years, a comprehensive and in-depths tutorial on the principles of SCMA from a beginner's point of view is missing, to the best of our knowledge. To promote this disruptive multiple access technique and to stimulate more forthcoming research activities on SCMA, we provide a self-contained introduction to SCMA (from codebook design to MPA decoding) in a systematic manner. In particular, we explain how efficient MPA decoding is carried out with the aid of a graphical model. 
\subsection{Organization of this work}
We start with the basic representation of an SCMA system and CB designing in Section II via the signature matrix. Section III is focused on the decoding using MPA. %In Subsection  3.1, an example on (3,1) repetition code with both hard decoding and soft decoding is discussed showing the advantages of the latter one. 
Specifically, in Subsections  III-A and III-B, maximum a posteriori (MAP) estimation and factor graphs are introduced, respectively. Subsection III-C presents a high-level introduction on the MPA, whilst subsection III-D demonstrates how belief messages are passed from function nodes to variable nodes and vice-versa in a factor graph. Subsection III-E presents the sum-product algorithm (SPA, a realization approach of MPA) carried out by passing belief messages over a factor graph with the aid of an example. Subsection III-F discusses the SPA and max-log-SPA used for multi-user detection of SCMA signals. Finally, the paper is concluded in Section IV.

\underline{\emph{Notations:}} In this paper, x, \textbf{x}, \textbf{X} denote a scalar, vector and matrix, respectively. Symbols $\textbf{x}^{T}$  and  $\textbf{X}^{T}$ represent transpose of \textbf{x} and \textbf{X}, respectively. Symbols $\mathbb{B}$ and $\mathbb{C}$ %, $\mathbb{Z}$ and $\mathbb{R}$
represents the set of binary numbers and complex numbers, %integers and real numbers, 
respectively. Also, the ${i}$th element of vector $\textbf{x}$ is denoted by $\textbf{x}_i$ and $\text{diag(\textbf{x})}$ denotes a diagonal matrix in which $i$th diagonal element is $\textbf{x}_i$.   `arg max[$f(x)$]' denotes value of $x$ that maximizes the function $f(x)$ and `log(\textbf{x})' denotes the natural logarithm of each element of $\textbf{x}$, respectively. `$\text{max}_x f(x)$' denotes the maximum value of $f(x)$ as $x$ is varied over all its possible values. $N \times M$ matrix denotes a matrix with $N$ rows and $M$ columns and $\textbf{I}_N$ denotes a  N × N identity matrix, respectively. 
\section{ System Model}
\subsection{SCMA Encoding and Codebook Design}

Consider an uplink synchronous SCMA system in which $J$ users transmit data to the basestation (BTS) using $K$ resource elements (e.g., time- or frequency- slots). In SCMA, the data/input bits of user is mapped to a complex codeword using the SCMA encoder. For instance, user $j$ wants to transmit $\textbf{b}_j$ bits. The encoder will map these $\textbf{b}_j$ bits to a codeword $\textbf{m}_j$ selected from a pre-defined codebook $\textbf{CB}_j$ as shown below
\begin{align}
 \textbf{m}_j= \textbf{CB}_j (\textbf{b}_j).   
\end{align}
where $\mathbf{m}_j \in \mathbb{A}_j \hspace{0.1cm} \subset \hspace{0.1cm} \mathbb{C}^K$, $\mathbb{A}_j$ denotes the set of codewords of $j$th user.

%Using a well designed codebook (CB) can significantly increase the spectrum efficiency as well as reduce the detection complexity. 
An SCMA encoder has J layers and there is a specific codebook (CB) dedicated to each user. Assuming one layer per user and from here on `user' and `layer' are used interchangeably. 
The performance gain of SCMA over other NOMA schemes is strongly dependent on well designed sparse codebooks.  The codebook of each user has its own sparsity pattern and can be written as a matrix of size $K\times M$, where $M$ denotes the number of codewords (i.e., columns of a CB matrix) allotted to a user. In a CB, each column vector (i.e., codeword) is sparse consisting of  $d_v$ non-zero elements at certain fixed resources elements (REs) pertinent to a specific user.

Albeit numerous SCMA CBs have been proposed in the literature, the optimal CB design remains an open challenge. The current designs are mostly sub-optimal which are based on a  multi-stage approach. For the $j$th user, multi-dimensional codebooks can be expressed as
\begin{align}\label{cb}
    \text{CB}_j=\textbf{V}_j \Delta_j \textbf{A}_{MC}^{'},~~ \text{for}~j=1,2,\cdots,J.
\end{align}
where $\textbf{V}_j \in $ \hspace{0.1cm} $\mathbb{B}^ {K \times d_v}$ denotes the binary mapping matrix, %of size $K\times d_v$
$\textbf{A}_{MC}^{'}$ denotes the multi-dimensional mother constellation   and $\Delta_j$ refers to the constellation operator for the $j$th user, respectively.
%\cite{cai2016multi}.  
%with the aid of multi-dimensional mother constellations, mapping matrices,  and user specific operations such as interleaving, permutations, and phase rotations. 
% Specifically, for  user $j$, bits $\textbf{b}_j$ are mapped to a complex codeword $\mathbf{m}_j$ using an SCMA encoder. 
%\begin{align}
    %    \textbf{m}_j = \textbf{V}_j (\Delta_j \circ \rho) \textbf{b}_j, ~~ \text{for}~ j=1,\cdots,J.
%\end{align}
%where $\mathbf{m}_j \in \mathbb{A} \hspace{0.1cm} \subset \hspace{0.1cm} \mathbb{C}^K$, $\mathbb{A}$ denotes the set of codewords allocated to user $j$, cardinality $|\mathbb{A}|=M$. $\textbf{V}_j$ \in \hspace{0.1cm} \mathbb{B}$^ {K \times N}$ denotes the binary mapping matrix of size $K\times N$, $\rho$ denotes the mother constellation   and $\Delta_j$ refers to the constellation operation for the $j$th user \cite{cai2016multi}.  
The mapping matrix is selected in such a way that each user has active transmissions over a few fixed REs only. 
The allotment of REs  among users can be represented by a signature matrix $\textbf{F}=[\textbf{f}_1,\textbf{f}_2,\cdots,\textbf{f}_J]$, where $\textbf{f}_j=\text{diag}(\textbf{V}_j \textbf{V}_j^T)$. 
$\textbf{F}_{jk}=1$ denotes that user $j$ has active transmission over the  $k$th RE. For a (4,6) SCMA block having $K=4$ REs and $J=6$ users, one example of the signature  matrix of size $4 \times 6$ is given as  %\cite{yu2018design}
\begin{align}\label{fac_gra_4_6}
    \textbf{F}_{4 \times 6}=
    \begin{bmatrix}
    1&1&1&0&0&0\\
    1&0&0&1&1&0\\
    0&1&0&1&0&1\\
    0&0&1&0&1&1\\
    %0&1&0&1&0&1&1&0\\
    %0&0&1&0&1&1&0&1\\
    \end{bmatrix}.
\end{align}
The binary mapping matrices corresponding to the six users are given below
\begin{equation*}
\begin{split}
    \textbf{V}_1 =
\begin{bmatrix}
1 & 0\\
0&1\\
0 & 0\\
0 &0\\
\end{bmatrix}, 
\textbf{V}_2 =
\begin{bmatrix}
1 & 0\\
0 & 0\\
0 & 1\\
0 & 0\\
\end{bmatrix}, 
\textbf{V}_3 =
\begin{bmatrix}
1 & 0\\
0 & 0\\
0 & 0\\
0 & 1\\
\end{bmatrix}, 
\\\\
\textbf{V}_4 =
\begin{bmatrix}
0 & 0\\
1 & 0\\
0&1\\
0 &0\\
\end{bmatrix},
 \textbf{V}_5 =
\begin{bmatrix}
0 & 0\\
1 & 0\\
0 & 0\\
0&1\\
\end{bmatrix},
\textbf{V}_6 =
\begin{bmatrix}
 0&0\\
0 & 0\\
1 &0\\
0&1\\
\end{bmatrix}.
\end{split}
\end{equation*}
$\textbf{V}_1$ indicates that user 1's data are transmitted over the first, and the second REs. Similarly, $\textbf{V}_2$ indicates that user 2's data are transmitted over the first and the  third REs, and so on. Once the mapping matrices are decided, one also requires mother constellation $\textbf{A}_{MC}'$ and layer-specific operations, $\Delta_j$. %Such a mother constellation are complex multi-dimensional vector sets having maximum minimum Euclidean distance or maximum minimum product distance.
Unitary rotations may be applied to a mother constellation to increase power variation among different users in order to reinforce the ``near-far effect" for suppression/mitigation of multiuser interference as well as to enhance the constellation shaping gain. Once the mother constellation is designed, layer-specific operations are applied to generate multiple CBs for different users.  These operations may include phase rotation, complex conjugate, layer power offset, and dimensional permutation.
%It is generally observed that higher the power diversity among interfering users, it becomes easy to decode interfering codewords at the receiver. 

\begin{figure*}[htbp]
\centering
\includegraphics[scale=0.4,trim=1cm 1cm 0cm 0.5cm,clip]{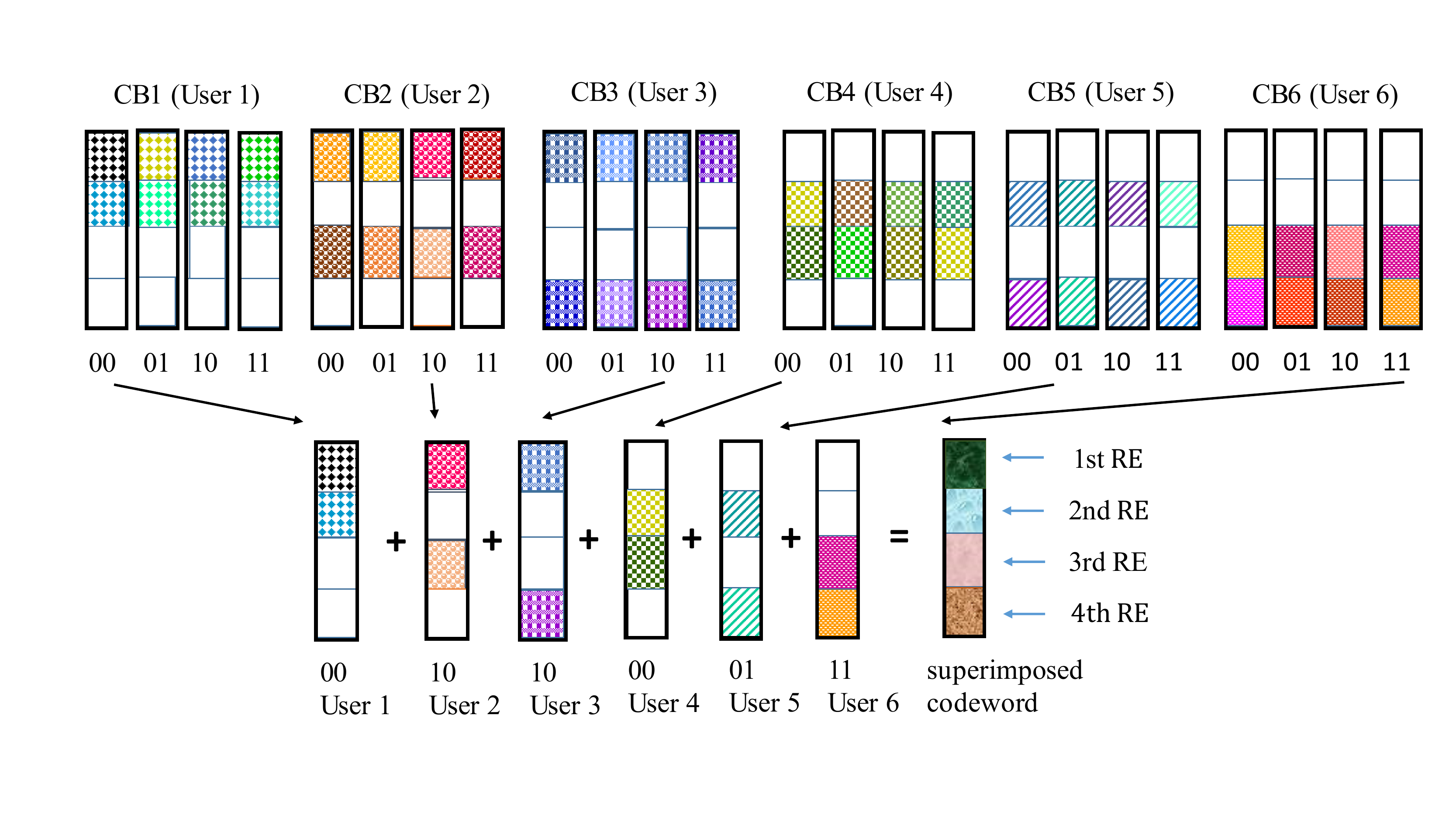}\\
\caption{ An illustration of $4\times 6$ SCMA  Encoding.}
\label{fig:encode}
\end{figure*}

 We provide an example of SCMA codebook design for $J=6, K=4$ \cite{interleaving} (therefore with overloading factor $J/K=1.5$) and the number of users superimposed over each RE is $d_f=3$. Steps 1-3 involve designing of mother codebook and Step 4 discusses user-specific operations.\\
$-$    \emph{Example 1:}
\begin{itemize}
    
\item{Step 1:} Initially a complex  vector $\mathbf{\Lambda}_1$ is designed.\\
    $\mathbf{\Lambda}_1=\{Y_m(1+i) \vert Y_m=2m-1-M,~m=1,\cdots,M\}$.\\
For $M=4,$ $\mathbf{\Lambda}_1=$  $\{-3(1+i),-1(1+i),1(1+i),3(1+i)\}$.
\item{Step 2:} For $d_v$ non-zero elements in a codeword,
    $\mathbf{\Lambda}_{d_v}=\textbf{U}_{d_v} \mathbf{\Lambda}_1$, where $ \textbf{U}_{d_v}=\text{diag}(\textbf{1} e^{i \theta_{d_v}})$ is a phase rotation matrix, $\textbf{1}$ is a $d_v$-dimensional vector of all ones and 
    $\theta_{d_v}= \frac{(d_v-1)\pi}{{d_v}}$.  Such an $d_v$-dimensional mother constellation is given below:
    \begin{multline*}
        \begin{split}
    \textbf{A}_{MC}&=[\mathbf{\Lambda}_1, \cdots,\mathbf{\Lambda}_{d_v}]^T \\
    &= \begin{bmatrix}
    \lambda_{11}& \lambda_{12} & \cdots & \lambda_{1M}\\
    \lambda_{21}& \lambda_{22} & \cdots & \lambda_{2M}\\
    \vdots & \vdots & \vdots & \vdots\\
    \lambda_{{d_v}1}& \lambda_{{d_v}2} & \cdots & \lambda_{{d_v}M}\\
        \end{bmatrix}
    {.}
    \end{split}
    \end{multline*}
     \item {Step 3:}  \emph{Interleaving}:  The elements of even dimensions of mother constellation $\textbf{A}_{MC}$ are reordered, as interleaved codewords performs good in fading channel and PAPR of mother constellation also reduces with interleaving. For $d_v=2, M=4,$ elements of second row $\mathbf{\Lambda}_2=\{\lambda_{21},\lambda_{22},\lambda_{23},\lambda_{24}\}$ can be re-ordered as \\
    ${\mathbf{\Lambda}_{2}^{'}}=\{\lambda_{22}, \lambda_{24},\lambda_{21}, \lambda_{23}\}$.\\
    The final mother constellation is ${\textbf{A}_{MC}^{'}}=[\mathbf{\Lambda}_{1},{\mathbf{\Lambda}_{2}^{'}},\mathbf{\Lambda}_{3}]^T.$
    
    \item {Step 4:} \emph{Generating codebook for different users.}\\ 
    Once the mother constellation is ready, the SCMA codebook is generated for each user. 
    The phase rotation angle $\omega_u$ can be given as
    \begin{align*}
        \omega_u= \text{exp}\left( \frac{i\pi u}{ d_v d_f} \right),~ \text{for}~ u=0,1,\cdots,d_f-1.
    \end{align*}
    These $\omega_u$ can be assigned to the non-zero positions of the factor graph matrix by Latin order. For $d_v=2,$ $M=4,$ $d_f=3$ and $J=6$ users, we have \\
    $\omega_0=1,$ $ \omega_1=\text{exp}\left( \frac{i \pi}{6} \right),$ $ \omega_2=\text{exp}\left( \frac{i \pi}{3} \right)$. \\
    The signature matrix with phase rotation angles is  given by
    \begin{align*}
        \mathbb{F}=
    \begin{bmatrix}
    \omega_{0}& \omega_{1} &\omega_{2}&0&0&0\\
    \omega_{1}&0&0&\omega_{2}&\omega_{0}&0\\
    0&\omega_{2}&0&\omega_{0}&0&\omega_{1}\\
    0&0&\omega_{0}&0&  \omega_{1}&\omega_{2}
        \end{bmatrix}
    {.}
    \end{align*}
    The operator for the $j$th user is  $\Delta_j=\text{diag}(\textbf{w}_j)$, where $ \textbf{w}_j$ is the $j$th column of $\mathbb{F}$ with only non-zero values.\\
    With mother constellation ${\textbf{A}_{MC}^{'}}$, constellation operator $\Delta_j$ and  binary  mapping  matrix $\textbf{V}_j$ for the $j$th user, codebook for the $j$th user can be generated using (\ref{cb}).
\end{itemize}
%An example of codebook for (4,6) SCMA block using which 6 users can transmit their data using 4 REs is given in \cite{klimentyev2016low}.

\subsection{System Model}
 Consider a symbol-synchronous uplink SCMA system where $J$ users communicate over  $K$ resource elements (REs). Let $\textbf{m}_j$ be the transmitted codeword of the $j$th user, where $\textbf{m}_j \in \mathbb{C}^{K \times 1}$ has cardinality $M= 2^{b}$ with $b$ denoting the number of bits per codeword.
Let us consider a (4,6) SCMA system with $M=4$, i.e., each codebook has 4 codewords that are mapped to two binary bits. 

 For an uplink SCMA system, let  $\textbf{h}_j =[h_{1j}, \cdots, h_{Kj}]^T$ be the effective channel fading coefficient for the $j$th user, where $h_{kj}$ denotes the channel fading coefficient at the $k$th RE for the $j$th user. Let  $\textbf{m}_j=[C_{1,j},\cdots,C_{K,j}]^T$ be the transmitted codeword of the $j$th user, where $C_{k,j}$ be the codeword element transmitted by the $j$th user on the $k$th RE. The received signal at the Base station is 
\begin{align}
    \textbf{y} = \sum_{j=1}^{J} \text{diag} (\textbf{h}_{j}) \textbf{m}_{j} + \textbf{n}.
    \end{align}
where $\textbf{n}$ $\in$ $\mathbb{C}^{K \times 1}$ is the noise vector, each element of which can be modeled as complex Gaussian distribution $\mathcal{C}\mathcal{N} [0,\sigma^2]$. 
Due to the sparse nature of SCMA codebooks, non-zero values from $d_f$ out of $J$ number of users overlap over each RE and also each user data is transmitted on $d_v<K$ REs.  Let $\xi_k$ be the set of users transmitting data over the $k$th RE and $\zeta_j$ be the set of REs on which the $j$th user has active transmission, respectively. By definition, the numbers  of elements in  $\xi_k$ and $\zeta_j$ are $d_f$ and $d_v$, respectively. Thus, the received signal at the $k$th RE is   
\begin{align}
        % y_k=\sum_{j_i \in N(k)}^{} h_{kj_i} C_{k,j_i}(\textbf{m}_{j_i}) +n_k, \hspace{1cm}     \text{for}~k=1,2,\cdots, K.
        %
        y_k=\sum_{j \in \xi_k}^{} h_{kj} C_{k,j}(\textbf{m}_{j}) +n_k, \hspace{1cm}     \text{for}~k=1,2,\cdots, K.
\end{align}
% where  $C_{k,j}(\textbf{m}_{j})$ denotes the codeword element  transmitted by user $j$ over the  $k$th RE.
 
$-$ \emph{Example 2:}
 Let us consider  (4,6) SCMA block encoding as shown in Fig. 1 which corresponds to the signature matrix $\textbf{F}_{4\times 6}$ (overloading factor $\lambda=J/K=1.5$) as shown in (\ref{fac_gra_4_6}). In $\textbf{F}_{4\times 6}$, each row corresponds a RE and each column corresponds to a user, respectively.
 \begin{comment}
 
 \begin{align}\label{fac_gra_4_6}
    \textbf{F}_{4 \times 6}=
    \begin{bmatrix}
    1&1&1&0&0&0\\
    1&0&0&1&1&0\\
    0&1&0&1&0&1\\
    0&0&1&0&1&1\\
    \end{bmatrix},
\end{align}
\end{comment}
 Note that the number of users superimposing over one  RE is 3, i.e., $d_f=3$ and the number of non-zero values in a column is 2, i.e., $d_v=2$. The received signal at the $k$th RE is
\begin{multline}
            y_k  =h_{k1}  C_{k,1} (\textbf{m}_1)+h_{k2}  C_{k,2} (\textbf{m}_2 )+h_{k3}  C_{k,3} (\textbf{m}_3 )+n_k,\\  ~~~~~~~~~~~~~~~~~~~~~~~~~\text{for}~k=1,2,\cdots, K.
\end{multline}
where $\textbf{m}_1,\textbf{m}_2,\textbf{m}_3$ are the codewords of users which belong to set $\xi_k$.  From $\textbf{F}_{4\times 6}$, the set of users which  interfere over the first  RE  is $\xi_1=\{1,2,3\}$ and the set of REs at which the first user has active transmission  is $\zeta_1=\{1,2\}$. Similarly,  $\xi_2=\{1,4,5\}$ implies that the first, the fourth, and the fifth users  superimpose at the second RE and $\textbf{m}_1,\textbf{m}_2,\textbf{m}_3$ are the codewords sent by users of set $\xi_2$.
%\fbox{\includegraphics[scale=0.5,trim=1cm 1cm cm 0.5cm,clip]{scma_encoding_factor_graph.pdf}}
% trim=left bottom right top
%\begin{figure}
%\centering
%\includegraphics[scale=0.4,trim=1cm 1cm cm 0.5cm,clip]{scma_encoding_factor_graph.pdf}
%\caption{ An illustration of $4\times 6$ SCMA  Encoding.}
%\label{fig:encode}
%\end{figure}
\section{Decoding}
The process of extracting  data from the noise corrupted received 
signal  is called decoding and there are two types of decoding, namely hard and soft decoding.
%\subsection{Soft Decoding}
In some communication systems, conventional hard decoding is  used, where each decision is taken at the receiver based on a certain threshold value. For instance, a simple hard decoding receiver decides that if  the received value is greater than the threshold value, it will be decoded as 1, and 0 otherwise. However, it is not verified how close or far is the received value to the threshold before making the decision. An alternative approach is to obtain the estimated sequence as well as its reliability level where the latter indicates the `confidence' we have in that estimated sequence. Such a decoding approach is called soft decoding \cite{book_soft_iter}. In this way, more information can be extracted from the received data and estimation/detection can be carried out in an iterative and  improved manner. We now present the following example to illustrate soft decoding. 
\begin{itemize}
    \item[$-$] \emph{Example 3:} (3,1) Repetition code:
\end{itemize}
%pdfcrop repetition_example_colourful.pdf
%\fbox{\includegraphics[scale=0.4,trim=.5cm 5cm 1.5cm 6.75cm,clip]{present.pdf}}
%trim=left bottom right top

\begin{figure*}[htbp]
\centering
\includegraphics[scale=0.4,trim=.5cm 5cm 1.5cm 6.75cm,clip]{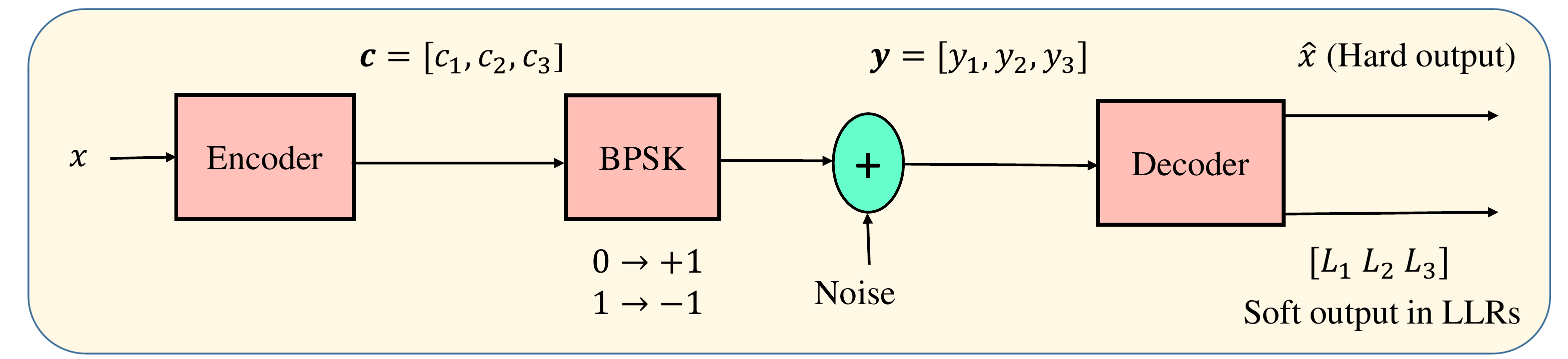}\\
\caption{(3,1) Repetition Code.}
\label{fig:rep}
\end{figure*}
Let $x$ be the bit to be transmitted. It is passed through the encoder which results in a codeword $\textbf{c}=[c_1, c_2, c_3]=[x, x, x]$ since it is a repetition code. After BPSK modulation, each $``0"$ bit is converted to ``1" and ``1" to $``-1"$, respectively. These modulated symbols propagate through a noisy channel, and the received signal can be expressed as $\textbf{y}=[y_1,y_2,y_3]$  as shown in Fig. \ref{fig:rep}. Next we discuss both hard decoding and soft decoding to recover the transmitted bit and understand the advantages of the latter.
\begin{enumerate}
    \item \textbf{Hard Decoding}\\
    Once $\textbf{y}$ is received at the receiver, every $y_i \in \textbf{y}$ goes through a threshold stage as shown in Fig. \ref{fig:hard}. If $y_i >0$,  $d_i=0$ otherwise $d_i=1$, where $d_i \in \textbf{d}$.  
 Let us assume that the received vector is  $\textbf{y}=[0.02, -2,-0.4]$. Using 0 as the hard threshold, $y_1=$ 0.02 is decided as $d_1= 0$ even though it is quite close to the threshold. Obviously, the confidence in this decision is very low. For $y_2=-2$, based on the threshold $d_2=1$ is decided and since it is far from the threshold, the confidence in this decision is large. Similarly, for $y_3=-0.4$, $d_3=1$ is decided. It is noted that the confidence level for $y_3$ is not as large as that for $y_2$, albeit larger than that for $y_1$. Thus, even though the bit sequence $\textbf{d}$ is estimated, still there is a lack of confidence that our decisions are correct.
 
 %\fbox{\includegraphics[scale=0.5,trim=2cm 11.5cm 4cm 3cm,clip]{hard_dec.pdf}}
% trim=left bottom right top
\begin{figure*}[htbp]
\centering
\includegraphics[scale=0.5,trim=2cm 11.5cm 4cm 3cm,clip]{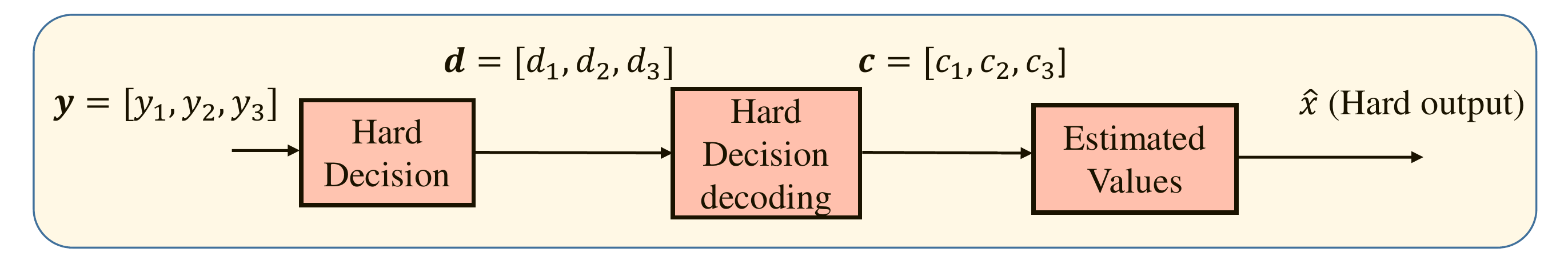}\\
\caption{Illustration of the hard decoding process.}
\label{fig:hard}
\end{figure*} 
 Once  the estimated bit sequence $\textbf{d}$ is determined, the Hamming distance for all possibilities for the  codeword  $\textbf{d}$ and $\textbf{c}$ is calculated. Since it's repetition code, $\textbf{c}=[x,x,x]$ can be either [000] or [111].  Then, $\textbf{c}$ with lower Hamming distance is chosen corresponding to every $\textbf{d}$. As we see in Table 1, first four values of 1st column are closer to $\textbf{c}=[0 0 0]$ and last four values are closer to $\textbf{c}=[1 1 1]$. Next, since its a repetition code, $\hat{x}$ is decided from $\textbf{c}$.
 %Table \ref{tab:kysymys}
  \begin{table}[]
 \begin{center}
 \begin{tabular}{||c c c ||} 
 \hline
 \textbf{d} & \textbf{c} & $\hat{x}$  \\ 
 \hline \hline
 000 & 000 & 0 \\ 
 \hline
 001 & 000 & 0 \\
 \hline
 010 & 000 & 0 \\
 \hline
 100 & 000 & 0 \\
 \hline
 011 & 111 & 1 \\ 
 \hline
 101 &111&1\\
 \hline
 110&111&1\\
 \hline
 111&111&1\\
 \hline
\end{tabular}
%\label{tab:kysymys}
\caption{Hard Decoding.}
\end{center}
 \end{table}
 
 %Hard decoding is a sub-optimal decoding.
With the repetition code, the same information is  transmitted multiple times. However, that information is not being efficiently utilized in hard decoding for the correct estimation.
 
\item {\textbf{Soft Decoding}}\\
For binary variables, the output vector is formed by three log-likelihood ratios (LLRs) $[L_1,  L_2, L_3]$ which are all real values indicating the ‘confidence or reliability level’ that $x_i$ bit is 0.
The  reliability of each received value is computed in order to improve the detection performance for the transmit data.  Using Bayes' rule\footnote{PMF and PDF are used when the output will result in discrete random variables and continuous random variables, respectively. For the AWGN channel% [5]
, transmitted symbols take discrete values and because of addition of noise in transmitted symbols, received values are continuous random variables.}:
\begin{align}\label{x1=0}
        P(c_1=0 \vert y_1)=\frac{f(y_1 \vert c_1=0) P(c_1=0)}{f(y_1)}\,,
\end{align}

\begin{align}\label{x1=1}
        P(c_1=1 \vert y_1)=\frac{f(y_1 \vert c_1=1) P(c_1=1)}{f(y_1)}\,.
\end{align}

Here, $P(c_1=0 \vert y_1)$ indicates the probability that $c_1=0$ given $y_1$ and $P(c_1=1 \vert y_1)$ indicates the probability that $c_1=1$ given $y_1$, respectively. $P(c_1=0)$ and $P(c_1=1)$ indicates the prior probabilities of the transmit bit, $f(y_1 \vert c_1=0)$ indicates the conditional distribution of $y_1$ given $c_1=0$, $f(y_1 \vert c_1=1)$ indicates the conditional distribution of $y_1$ given $c_1=1$ and $f(y_1)$ indicates the probability distribution function of $y_1$, respectively. By assuming BPSK modulation with equiprobable transmission, we have $P(c_1=0)=P(c_1=1)=1/2$.  Using  (\ref{x1=0}-\ref{x1=1}),
\begin{align}\label{ratio}
        \frac {P(c_1=0 \vert y_1)}{P(c_1=1 \vert y_1)}=\frac{f(y_1 \vert c_1=0) }{f(y_1 \vert c_1=1)} \,.
\end{align}
For $c_1=0$,  $y_1=1+\mathcal{N}(0,\sigma^{2})$ and for $c_1= 1$,  $y_1=-1+ \mathcal{N}(0,\sigma^{2})$. Thus,  (\ref{ratio}) can be calculated as 
\begin{multline}
\begin{split}
        \frac {P(c_1=0 \vert y_1)}{P(c_1=1 \vert y_1)}&=\frac{\frac{1} {{\sigma \sqrt {2 \pi}}} \exp \left({{{- \left ({y_1 - 1} \right) ^ 2} \mathord {\left / {\vphantom {{- \left ({x - \mu} \right) ^ 2} {2 \sigma ^ 2}}} \right. \kern- \nulldelimiterspace} {2 \sigma ^ 2}}}\right)}{\frac{1} {{\sigma \sqrt {2 \pi}}} \exp \left({{{- \left ({y_1 + 1} \right) ^ 2} \mathord {\left / {\vphantom {{- \left ({x - \mu} \right) ^ 2} {2 \sigma ^ 2}}} \right. \kern- \nulldelimiterspace} {2 \sigma ^ 2}}}\right)}\\
        &= \exp{\left(\frac{2 y_1}{\sigma^{2}}\right)} \,.
\end{split}
\end{multline}
The above calculation gives the likelihood ratio (or reliability) computed based on $y_1$ alone. By recalling the principle of repeated coding, every bit is repeated three times, meaning that the reliability of the transmit bit can be improved based on the three received values $y_1,y_2,y_3$. Therefore, the LLR of the transmit bit can be expressed as 
\begin{align}\label{llr_rep}
        L_1=\log {\frac{P( c_1=0 \vert y_1, y_2 ,y_3) }{P( c_1=1 \vert y_1, y_2 ,y_3)}}\,.
\end{align}
Again using Bayes’ rule,
\begin{align}\nonumber
        P(c_1=0 \vert y_1,y_2&,y_3)\\=&\frac{f(y_1,y_2,y_3 \vert c_1=0) P(c_1=0)}{f(y_1,y_2,y_3)}\,.
\end{align}

\begin{align}\nonumber
        P(c_1=1 \vert y_1,y_2,&y_3)\\=&\frac{f(y_1,y_2,y_3 \vert c_1=1) P(c_1=1)}{f(y_1,y_2,y_3)}\,.
\end{align}
Therefore,  (\ref{llr_rep}) can be simplified as
\begin{equation}
\begin{split}
L_1 &=\log {\frac{f(  y_1, y_2 ,y_3  \vert c_1=0) }{f(  y_1, y_2 ,y_3  \vert c_1=1)}}\
\\
  = \log&  \left(\frac{\exp \frac{-(y_1-1)^2}{2\sigma^2}}{\exp \frac{-(y_1+1)^2}{2\sigma^2}} 
        \frac{\exp \frac{-(y_2-1)^2}{2\sigma^2}}{\exp \frac{-(y_2+1)^2}{2\sigma^2}}
        \frac{\exp \frac{-(y_3-1)^2}{2\sigma^2}}{\exp \frac{-(y_3+1)^2}{2\sigma^2}}\right)\\ & =\frac {y_1+y_2+y_3} {2 \sigma^{2}}\,.
 %f_1 (a) f_2 (b)\,.
\end{split}
\end{equation}
\begin{comment}

\begin{align}
L_1=\log {\frac{f(  y_1, y_2 ,y_3  \vert c_1=0) }{f(  y_1, y_2 ,y_3  \vert c_1=1)}}\,,
\end{align}
        
\begin{align}
            = \log  \left(\frac{\exp \frac{-(y_1-1)^2}{2\sigma^2}}{\exp \frac{-(y_1+1)^2}{2\sigma^2}} 
        \frac{\exp \frac{-(y_2-1)^2}{2\sigma^2}}{\exp \frac{-(y_2+1)^2}{2\sigma^2}}
        \frac{\exp \frac{-(y_3-1)^2}{2\sigma^2}}{\exp \frac{-(y_3+1)^2}{2\sigma^2}}\right) \\
        =\frac {y_1+y_2+y_3} {2 \sigma^{2}}\,.
\end{align}
\end{comment}
Similar calculations can be done for $L_2$ and $L_3$.
Here, $\frac{1}{2\sigma^2}$ is a constant and can be omitted for simplicity. So, if $y_1+y_2+y_3 >0$, then $\hat{\textbf{c}}=[000]$ is decided otherwise $\hat{\textbf{c}}=[111]$. With soft decoding, information from all the received values is efficiently utilized. This way transmitted bits are estimated along with a confidence level.
\end{enumerate}

\subsection{MAP Detection}
The objective of an optimal detector is to minimize the probability of error ($P(e)$) for the transmitted bit sequence $\textbf{x}$ , i.e., to minimize the mismatch between transmitted bits ($\textbf{x}$) and estimated bits ($\hat{\textbf{x}}$).
\begin{align}\label{prob_err}
    \min P(e)= \min P(\textbf{x} \neq \hat{\textbf{x}})\,.
\end{align}
Given all the possible observations  \textbf{y} and the channel coefficient matrix \textbf{H}, we have
\begin{align}\label{error}
    P(e)= \int P(e \vert \textbf{y}) f(\textbf{y}) d\textbf{y}\,,
\end{align}
where $P(e \vert \textbf{y})$ indicates the probability of error given received vector $\textbf{y}$ and $f(\textbf{y})$ indicates the probability density function (pdf) of \textbf{y}, respectively. It is noted from (\ref{error})  that $P(⁡e)$ is proportional to $P(e \vert \textbf{y})$. 
Since  $\textbf{x}$ is the desired signal, we have
\begin{align}
    P(e \vert \textbf{y})=1-P(\textbf{x} \vert \textbf{y})\,,
\end{align}
where $P(\textbf{x} \vert \textbf{y})$ denotes the probability of $\textbf{x}$ given the vector $\textbf{y}$. Thus, to minimize the probability of error, $P(\textbf{x} \vert \textbf{y}) $ needs to be maximized and this detection scheme is termed as maximum a-posteriori (MAP) detection. Using Bayes’ rule,
\begin{align}
    P( \textbf{x} \vert \textbf{y})= \frac{f(\textbf{y} \vert \textbf{x})P(\textbf{x})}  {f \textbf{(y})}\,,
\end{align}
 where $f(\textbf{y} \vert \textbf{x})$ indicates the conditional pdf  of $\textbf{y}$ given $\textbf{x}$ and $f(\textbf{y})$ indicates the pdf of $\textbf{y}$, respectively. Since $f(\textbf{y})$ is constant for all the values that $\textbf{x}$ can take and assuming the prior probabilities of $P(\mathbf{x})$ are identical for different $\mathbf{x}$, so
 \begin{align}
     P(\textbf{x} \vert \textbf{y}) \propto f(\textbf{y} \vert \textbf{x} )\,.
 \end{align}
 Thus, the $ \textbf{x} $ which  maximizes $f(\textbf{y} \vert \textbf{x} )$ is chosen to minimize the probability of error and this detection scheme  is called the maximum likelihood (ML) detection. In short, the MAP detection is equivalent to the ML detection when prior probabilities take the identical value.\\\\
\textbf{MAP Detection in SCMA:}\\
For an SCMA decoder, given \textbf{y} and the channel matrix \textbf{H} at the receiver, the detected multiuser  codeword $\hat{\textbf{X}}=[\textbf{m}_1,\textbf{m}_2,\cdots,\textbf{m}_J]$ is given as
\begin{align} \label{multiuser_codeword}
    \hat{\textbf{X}} = \argmax_{\textbf{m}_j \in \mathbb{A}_j,\forall j} p(\textbf{X} \vert \textbf{y})\,,
\end{align}
where $\textbf{m}_j$ represents the codeword transmitted by the $j${th} user and $ \mathbb{A}_j$ represents the set of codewords allotted to the $j$th user, i.e., the $j$th user's codebook. The transmitted codeword of each user can be estimated by maximizing
its a-posteriori probability mass function (pmf). This can be calculated by taking
the marginal of the joint a-posteriori pmf defined in (\ref{multiuser_codeword}). So, the objective is to detect the codeword transmitted by each user one by one.
\begin{enumerate}
    \item [$\circ$] \textbf{\emph{Marginalization:}} It is known that that if $\upsilon$ is a function of a number of variables, then $\upsilon$ with respect to a specific variable can be obtained by carrying out marginalization with respect to that variable.

\begin{itemize}
    \item [$-$] \emph{Example 4:} Let $\upsilon(\alpha_1,\alpha_2, \cdots, \alpha_j,\cdots, \alpha_J)$ be a function (also called a global function) of $J$ number of variables, then the marginalized function with respect to $\alpha_j$ variable is given as
\begin{align}
    \upsilon(\alpha_j)= \sum_{\alpha_1} \cdots \sum_{\alpha_{j-1}} \sum_{\alpha_{j+1}} \cdots \sum_{\alpha_J} \upsilon(\alpha_1, \cdots, \alpha_J)\,,
\end{align}
i.e., the summation with respect to all variables except for $\alpha_j$. It can be written   in a more compact form as
\begin{align}\label{marginalization}
    \upsilon(\alpha_j)= \sum_{{\sim}\alpha_j}  \upsilon(\alpha_1,\alpha_2, \cdots \alpha_j,\cdots \alpha_J)\,.
\end{align}
where $\sim\alpha_j$ denotes that the summation is performed with respect to all the variables of the global function except for $v_j$.

\end{itemize}
\end{enumerate}
Similar to (\ref{marginalization}), the marginalization with respect to $\textbf{m}_j$ in  $(\ref{multiuser_codeword})$ leads us to symbol by symbol detection for the  $j$th user which can be given as
\begin{align}
    \hat{\textbf{m}}_j = \argmax_{\textbf{m}_j \in \mathbb{A}_j} \sum_{{\sim} \textbf{m}_j} p(\textbf{X} \vert \textbf{y})\,.
    %  p is pmf function.
\end{align}
Again using Bayes’ rule, we have
\begin{align}\label{map}
    \hat{\textbf{m}}_j = \argmax_{m_j \in \mathbb{A}_j} \sum_{{\sim} \textbf{m}_j} f(\textbf{y} \vert \textbf{X})  P(\textbf{X})\,.
\end{align}
For SCMA system, assume that the  noise components corresponding to the $K$ REs are identically distributed and independent of each other. As a result, we have 
\begin{align}
    f(\textbf{y} \vert \textbf{X})= \prod_{k=1}^{K} f(y_k \vert \textbf{X})\,,
\end{align}
where $k=1,\cdots,K$ denotes the REs and $y_k$ is the received signal at the $k$th RE. Consequently,  (\ref{map}) becomes
\begin{align}\label{mpf}\nonumber
    \hat{\textbf{m}}_j = \argmax_{\textbf{m}_j \in \mathbb{A}_j} \sum_{{\sim}\textbf{m}_j} \left(P(\textbf{X}) \prod_{k=1}^{K} f(y_k \vert \textbf{X})\right)\, 
    \enspace \\\text{for}~ j=1, \cdots, J.
\end{align}
Solving the marginal  product of functions (MPF) problem in  (\ref{mpf}) with brute force will lead to prohibitively high  complexity and hence may be infeasible when number of users is large. Thanks to the sparsity of SCMA codebooks, we will show in the sequel that low-complexity message passing algorithm (MPA) can be leveraged with the aid of $\textbf{factor graph}$ to solve the MPF problem in  (\ref{mpf}) with near-optimal performance.  

\subsection{Factor Graphs}
A graph is composed of vertices (or nodes) and edges. Two nodes are connected with an edge when there is some relationship between them. Different types of graphs are used to model problems in areas such as computer science, biology, physics,  etc. A widely known graph for modelling  communication and signal processing problems is called a bipartite graph. In this graph, total nodes can be divided in two sets and no two nodes within a set are connected to each other. Fig. 4. shows a bipartite graph having two set of nodes, set A and set B. It is noted that no two nodes of a set are connected.

%\fbox{\includegraphics[scale=0.5,trim=8.5cm 3cm 12cm 6.5cm,clip]{bipartite_gr.pdf}}
%trim=left bottom right top

\begin{figure}[h!]
\centering
\includegraphics[scale=0.4,trim=8.5cm 3cm 12cm 6.5cm,clip]{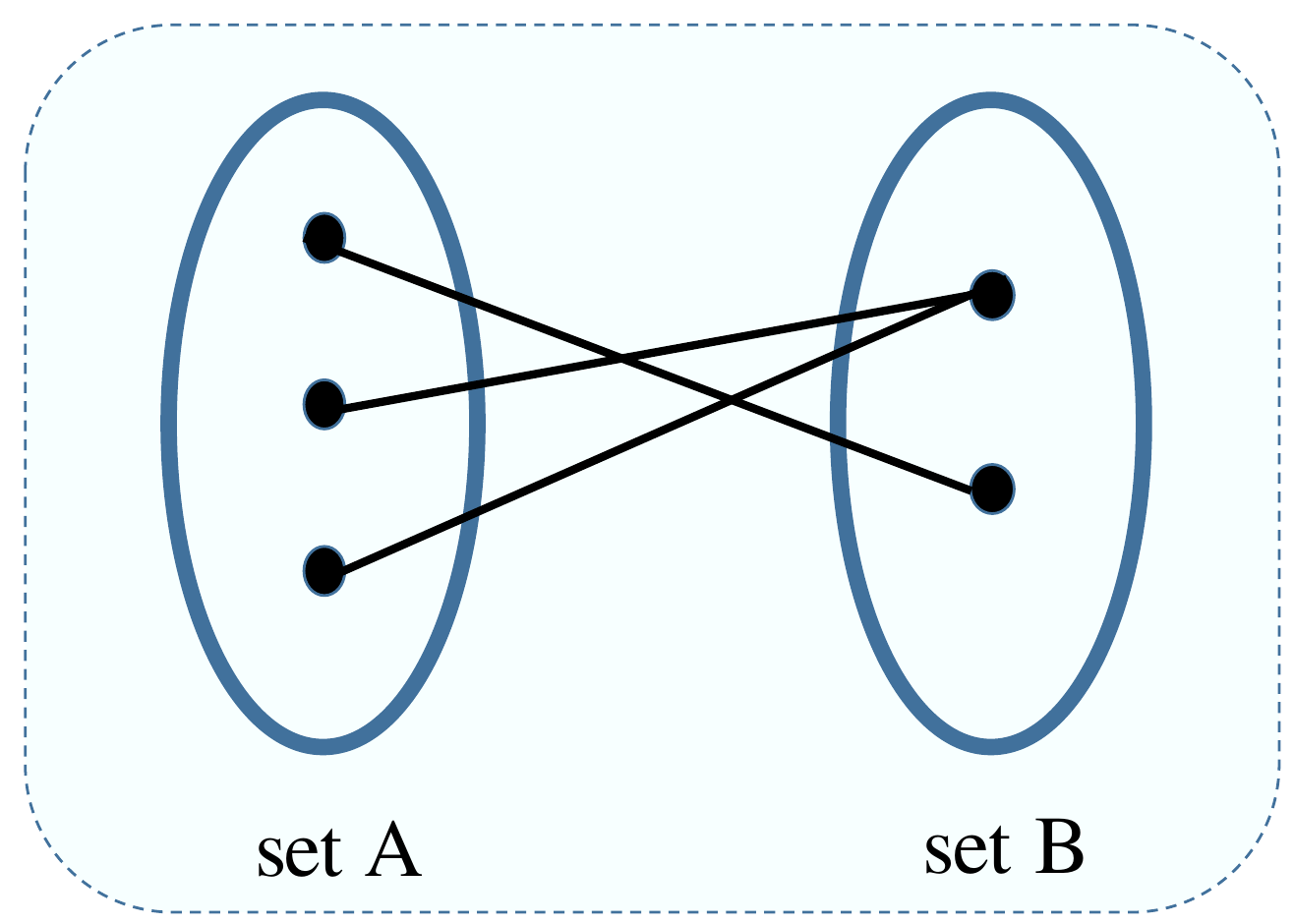}
\caption{Bipartite Graph.}
\label{fig:bipar_ex}
\end{figure}

A factor graph is an undirected bipartite graph in which one set of nodes is called variable nodes (VNs) and the other called function nodes (FNs). An edge is connected between a variable node and a function node if that particular variable is an argument of that function. Factor graph shows how a global function can be  represented in terms of simpler local functions (denoted by FNs) and can also help in computing marginal distribution with respect to single variable using sum-product algorithm (SPA). Next, some examples are illustrated on how a global function can be factorized and how factorization helps in reducing the computation complexity.
\begin{itemize}
    \item[$-$] \emph{Example 5:}  Let $\gamma$ be a global function of three variables ($a, b, c)$ such that $\gamma=a^2+a*b+a*c$. Let $a=1,b=3,c=4$, then it can be seen from Fig. 5(a) that function $\gamma$ can be computed in 11 steps (including inserting values into the variables). By rearranging  the global function to $\gamma=a(a+b+c)$, $\gamma$ can be computed in 7 steps (Fig. 5(b)). Thus, by simplifying a function, its computation complexity can be reduced.
%\fbox{\includegraphics[scale=0.5,trim=8cm 4cm 12cm 3cm,clip]{fact_ex2.pdf}}
%trim=left bottom right top

\begin{figure}[htbp]
    \centering
    \subfigure[]{
    \includegraphics[scale=0.3,,trim=8cm 4cm 11cm 2.5cm,clip]{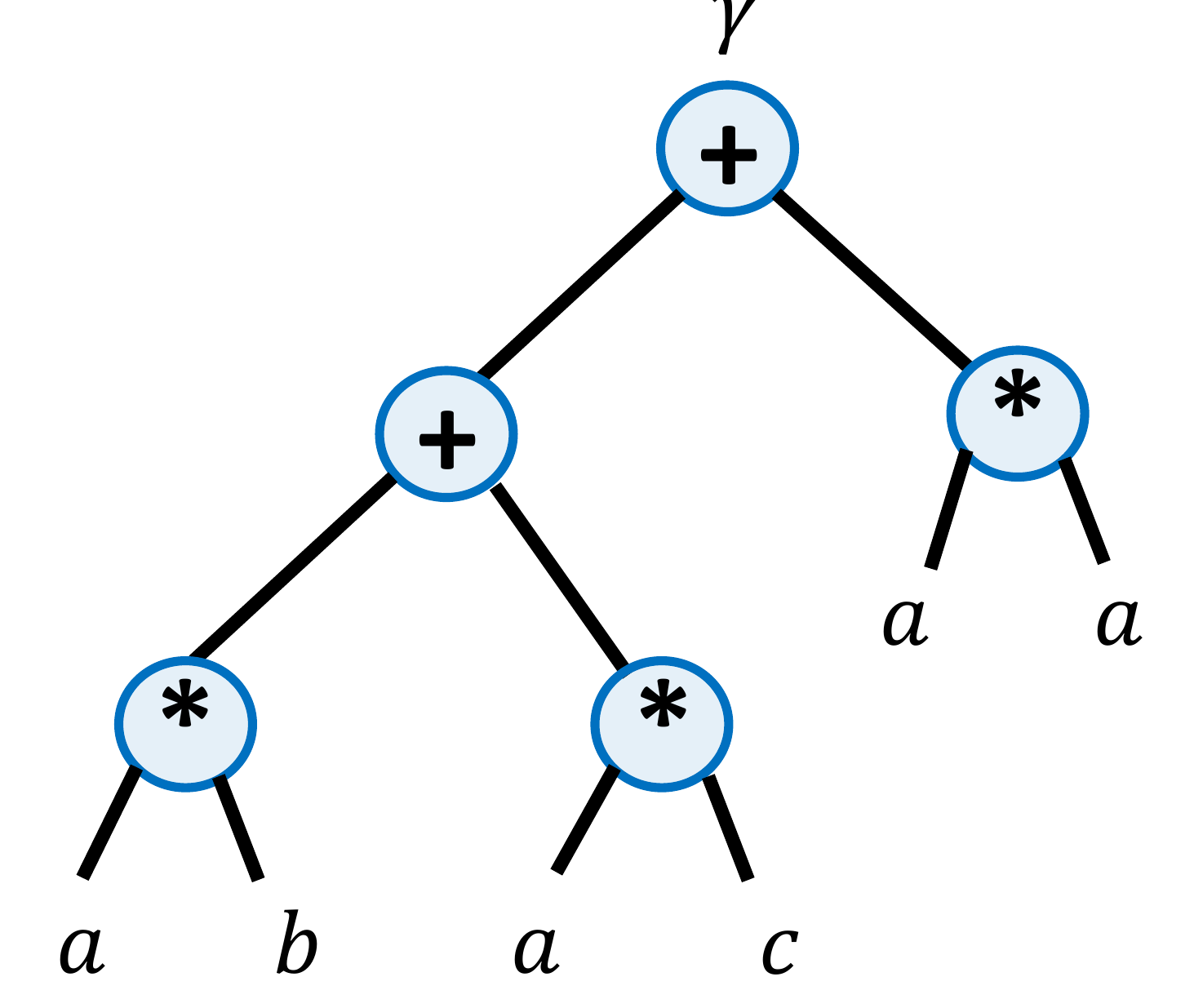}}

    \subfigure[]{
    \includegraphics[scale=0.3,trim=8cm 4cm 12cm 2cm,clip]{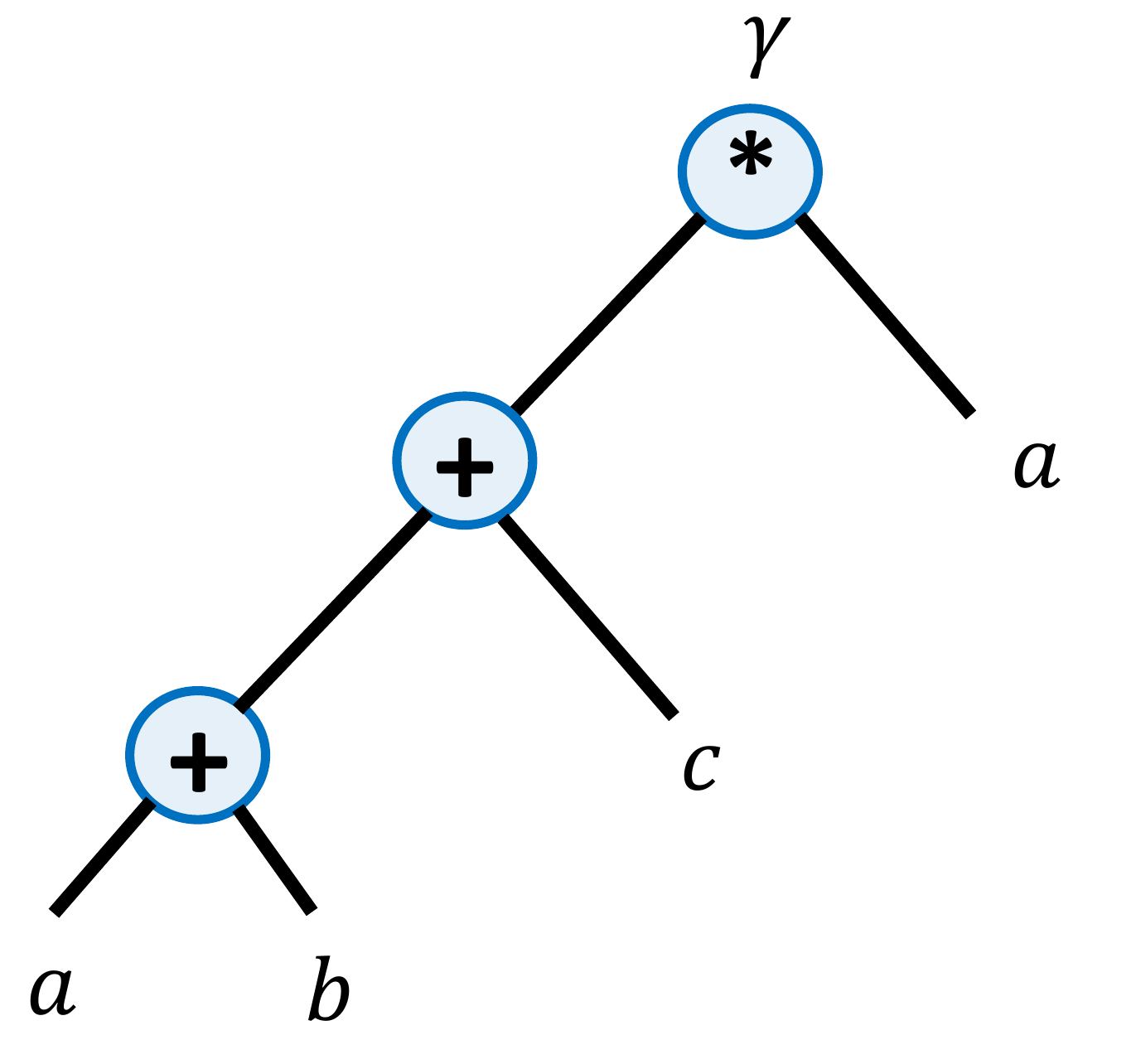}}
    
    \caption{Graphical model of (a) $\gamma=a^2+a*b+a*c$ (b) $\gamma=a(a+b+c)$.}
    \label{fig:fact_ex}
    \end{figure}

\item [$-$]  \emph{Example 6:} Joint distribution of two variables $(a, b)$ can also be simplified in conditional and prior probability distribution functions.
\begin{equation} \label{eq1}
\begin{split}
f(a,b) & =f_1(a \vert b)f_2(b)\ \\
 & = f_1' (a) f_2 (b)\,.
\end{split}
\end{equation}

\end{itemize}
Factor graphs are also widely used to model a number of communication problems. Fig. \ref{fig:fac_comm_ex} shows the factor graph of a communication channel.%~\cite{tarokh}.
%\fbox{\includegraphics[scale=0.4,trim=3cm 4.5cm 5cm 1.75cm,clip]{fact_comm_ex.pdf}}
%trim=left bottom right top
\begin{figure}[h!]
\centering
\includegraphics[scale=0.3,trim=3cm 4.5cm 5cm 1.75cm,clip]{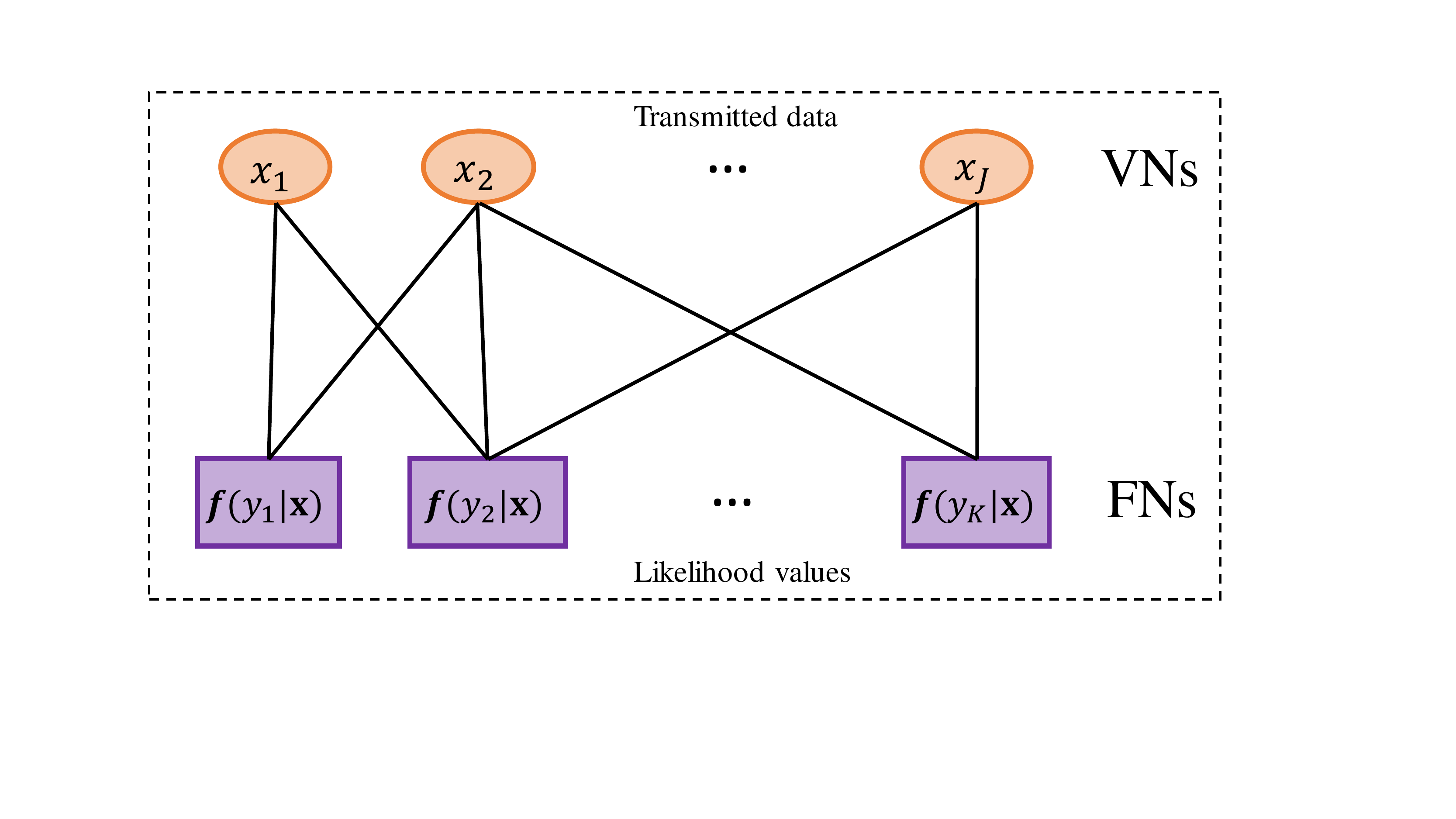}
\caption{Factor graph of a communication channel.}
\label{fig:fac_comm_ex}
\end{figure}
Let $\textbf{x}=[x_1,x_2,\cdots, x_j,\cdots,x_J]$ be the transmit  vector and $\textbf{y}=[y_1 , y_2, \cdots, y_k, \cdots,y_K]$ be the received vector, respectively. Assuming  noise components are independent of each other, then
\begin{equation}
    f(\textbf{y} \vert \textbf{x})=\prod_{k=1}^{K} f(y_k \vert \textbf{x})\,.
\end{equation}
It is noted that each received symbol ($y_k$) depends on certain elements of \textbf{x}. Let say $y_k$ depends on $x_1,x_2,x_j$ elements, then there will be edge connecting FN corresponding to $f(y_k \vert x_1,x_2,x_j)$ with VNs $x_1,x_2,x_j$. In Fig. \ref{fig:fac_comm_ex}, every VN denotes the data from one transmit point and every FN $k$ denotes the likelihood function $f(y_k \vert \textbf{x})$.

\subsection{Message Passing Algorithm}
\begin{figure*}[h!]
    \centering
    \subfigure[]{    \includegraphics[scale=0.45,trim=9.25cm 5.5cm 11.25cm 5.5cm,clip]{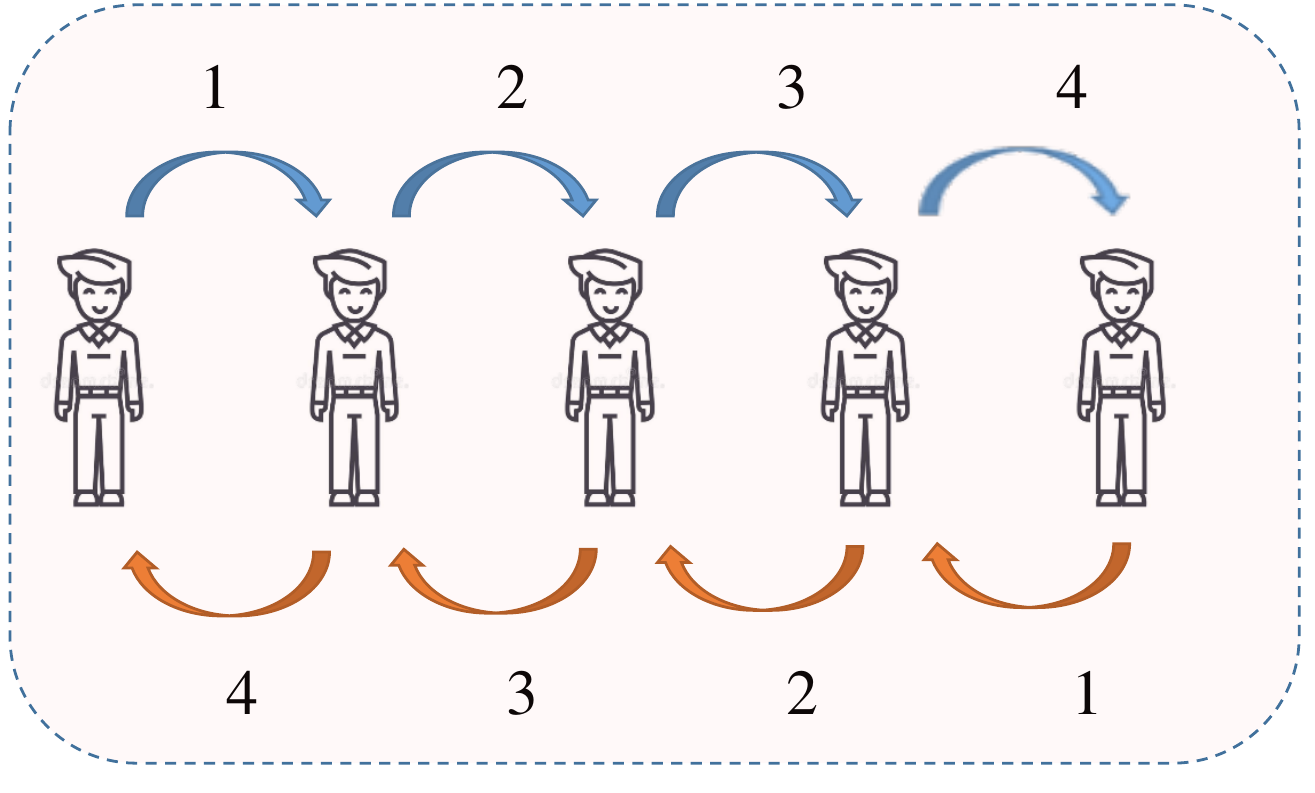}}
    %\label{fig:Mpa_ex_a}
    \subfigure[]{
    \includegraphics[scale=0.45,trim=9.25cm 5.5cm 11.25cm 5.25cm,clip]{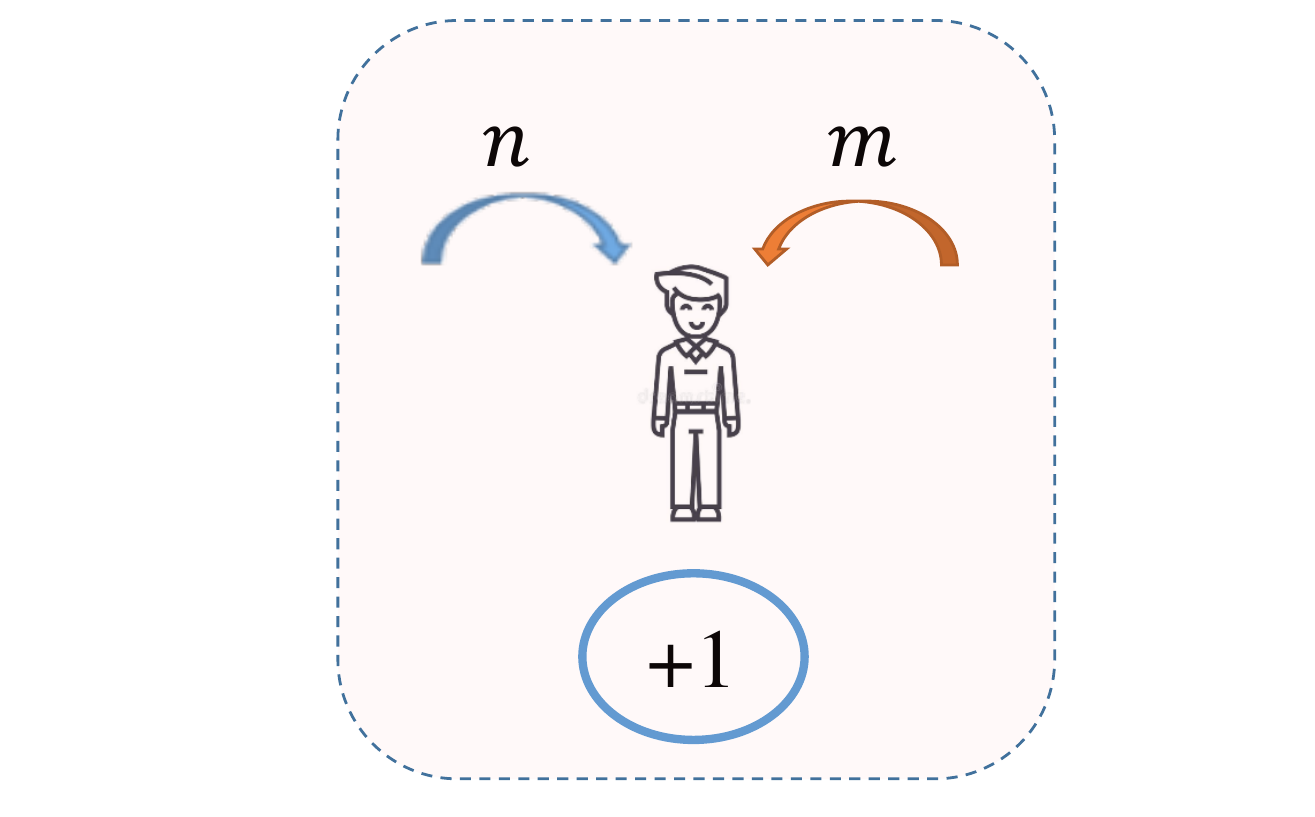}}
    %\label{fig:Mpa_ex_b}
    \caption{Students counting using MPA.}
    \label{fig:Mpa_ex}
    \end{figure*}

Message passing algorithm (MPA) is an algorithm to conduct inference from graphical models by passing belief messages between the nodes. The following example may be helpful in understanding how inference can be obtained from a graph \cite{book_soft_iter}.
\begin{itemize}
    \item[$-$] \emph{Example 7:} Fig. \ref{fig:Mpa_ex} (a) shows  a group of students standing in a line. The objective is to count total number of students and pass this information to each student. The constraint is that one student can communicate with maximum two  neighboring students (front and back) at a time. To solve this problem, the approach proposed by MPA is, that when one student receives a count from one side, he/she adds 1 to it (to indicate the student's presence) and pass it to the other side. The algorithm starts from each end of the line where the counter starts from 1 and increases by 1 as the message passes through each student. Messages pass in both directions of the line. If a student receives a count of $n$ from one side and $m$ from another side, then the total students present are $n+m+1$  (as shown in Fig. \ref{fig:Mpa_ex} (b)). 
    In this way, every student obtains the information of total number of students present in the line.
    Such type of problems can also be modelled with the help of a graph, where each node represents a student and every connecting edge represents a communication link between them.
\end{itemize}
%\fbox{\includegraphics[scale=0.5,trim=9.25cm 5.5cm 11.25cm 5.5cm,clip]{mpa_example_a.pdf}}
%trim=left bottom right top
%\fbox{\includegraphics[scale=0.5,trim=9.25cm 5.5cm 11.25cm 5.25cm,clip]{mpa_example_b.pdf}}
\subsection{Message Passing Between Nodes}
In this subsection, we introduce how messages are passed between function nodes (FNs) and variable nodes (VNs) of a factor graph using Sum-product algorithm (SPA) (a version of Message passing algorithm). In SCMA systems, each VN denotes one data layer and each FN denotes the likelihood function at the resource element (RE). Therefore, the total number of VNs is equal to the total number of layers/users and the total FNs equals the total REs present.

Suppose the transmitted bits are $\textbf{c}= [c_1, c_2 \cdots, c_N]$ and received bits are $\textbf{y}= [y_1,y_2, \cdots, y_N]$, then the  aim is to compute the \textit{a posteriori} probability (APP) of bit $c_i$, i.e., $P(c_i=0 \vert \textbf{y})\,.$
%\begin{align*}
    %P(c_i=0 \vert \textbf{y})\,.
%\end{align*}
Using Bayes' rule, the APP ratio with regard to $c_i$   can be converted into likelihood ratio as follows: 
\begin{align*}
    \frac{P(c_i=0 \vert \textbf{y})}{P(c_i=1 \vert \textbf{y})} = \frac{f(\textbf{y} \vert c_i=0)}{f(\textbf{y} \vert c_i=1)}\,.
\end{align*}
Taking natural logarithm, we obtain the Log-likelihood ratio (LLR) of $c_i$ below
\begin{align*}
    \text{LLR}(c_i)= \log \left(\frac{f(\textbf{y} \vert c_i=0)}{f(\textbf{y} \vert c_i=1)}\right)\,.
\end{align*}
If $\text{LLR}(c_i)<0$, then $c_i=1$ is decoded otherwise 0. \\
Message passing in a factor graph using SPA is an iterative process if the factor graph has cycles (closed loops) %\footnote{A cycle is a path which starts from a node, passes through the edges and ends at the starting node.}
 present in it.
In every iteration, there are two steps. In Step 1, a belief message is passed from a variable node (VN) to a function node (FN) and in Step 2, the message is passed from an FN to a VN, respectively. These two steps are discussed in detail as follows: 
\begin{itemize}
     \item {Step 1}: Suppose  there is a VN $j_1$ which has connections with 3 FNs with indices  $k_1,k_2,k_3$ as shown in Fig. \ref{fig:vn_fn}. To pass a  message from VN $j_1$ to FN $k_2$, firstly VN $j_1$ multiplies all the messages received from its neighboring nodes except FN $k_2$ (i.e., $k_1$ and $k_3$) and then transfer the output to FN $k_2$.
     \begin{align*}
         n_{j_1 \rightarrow k_2}= n_{k_1 \rightarrow j_1}  \enspace n_{k_3 \rightarrow j_1}.
     \end{align*}
     Here, $n_{j_1 \rightarrow k_2}$ indicates the belief message from VN $j_1$ to FN $k_2$.
     The outgoing message from a VN is in the form of either $P(c_i=0 \vert \textbf{y})$ or APP ratio or likelihood ratio.
  %   trim=left bottom right top
%\fbox{\includegraphics[scale=0.5,trim=11.5cm 5.5cm 13cm 4cm,clip]{vn_t_fn.pdf}}

     \begin{figure}[h!]
\centering
\includegraphics[scale=0.4,trim=11.5cm 5.5cm 13cm 4cm,clip]{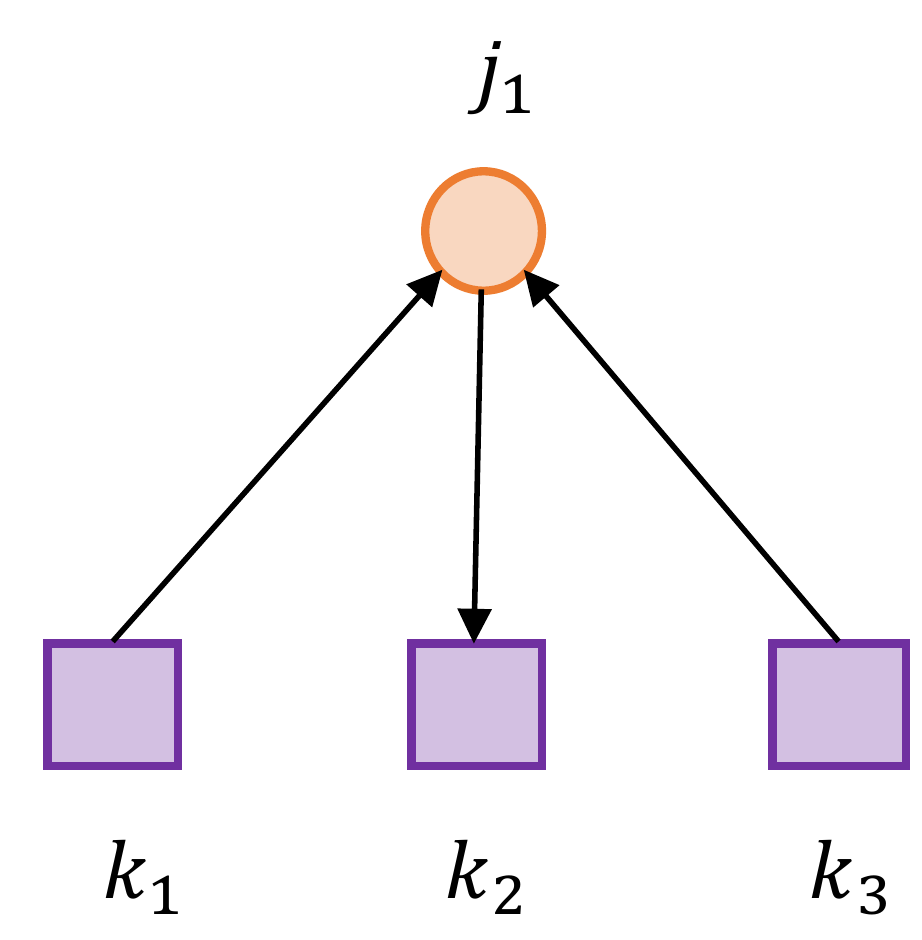}
\caption{Message passing from a VN to an FN.  }
\label{fig:vn_fn}
\end{figure}
     
     \item {Step 2}: In this step, a belief message is passed from an FN to a VN. Consider a FN $k_1$ which has three neighboring VNs ($j_1,j_2,j_3$) as shown in Fig. \ref{fig:fn_vn}. To send a belief message from FN $k_1$ to VN  $j_2$, FN $k_1$ first collects all the messages from its neighboring nodes except for VN $j_2$. These received messages are multiplied with the local function ($f_{k_1}(j_1,j_2,j_3)$) associated with  FN $k_1$ and then the resulting function is marginalized with respect to VN $j_2$. After marginalization, the resulting message to be sent to VN $j_2$ can be expressed as 
     
     \begin{align*}
         n_{k_1 \rightarrow j_2}= \sum_{{\sim}j_2} {( f_{k_1}(j_1,j_2,j_3) \enspace n_{j_1 \rightarrow k_1} \enspace n_{j_3 \rightarrow k_1})} .
     \end{align*}
     
     Here, $f_{k_1}(j_1,j_2,j_3)$ indicates the local function of FN $k_1$ and  message $n_{k_1 \rightarrow j_2}$ indicates $P$(function $k_1$ is satisfied $\vert$ messages received at FN $k_1$), respectively. Similarly, if a belief message needs to be passed from FN $k_1$ to VN $j_3$, then the belief message from the VNs  $j_1$ and $j_2$ (all VNs except  the one to which message needs to be passed) is considered as extrinsic information.%or ratio of these probabilities or log of ratio of such probabilities. %
       %   trim=left bottom right top
%\fbox{\includegraphics[scale=0.5,trim=11cm 5.5cm 13cm 4.5cm,clip]{fn_t_vn.pdf}}
\begin{figure}[h!]
\centering
\includegraphics[scale=0.4,trim=11cm 5.5cm 13cm 4.5cm,clip]{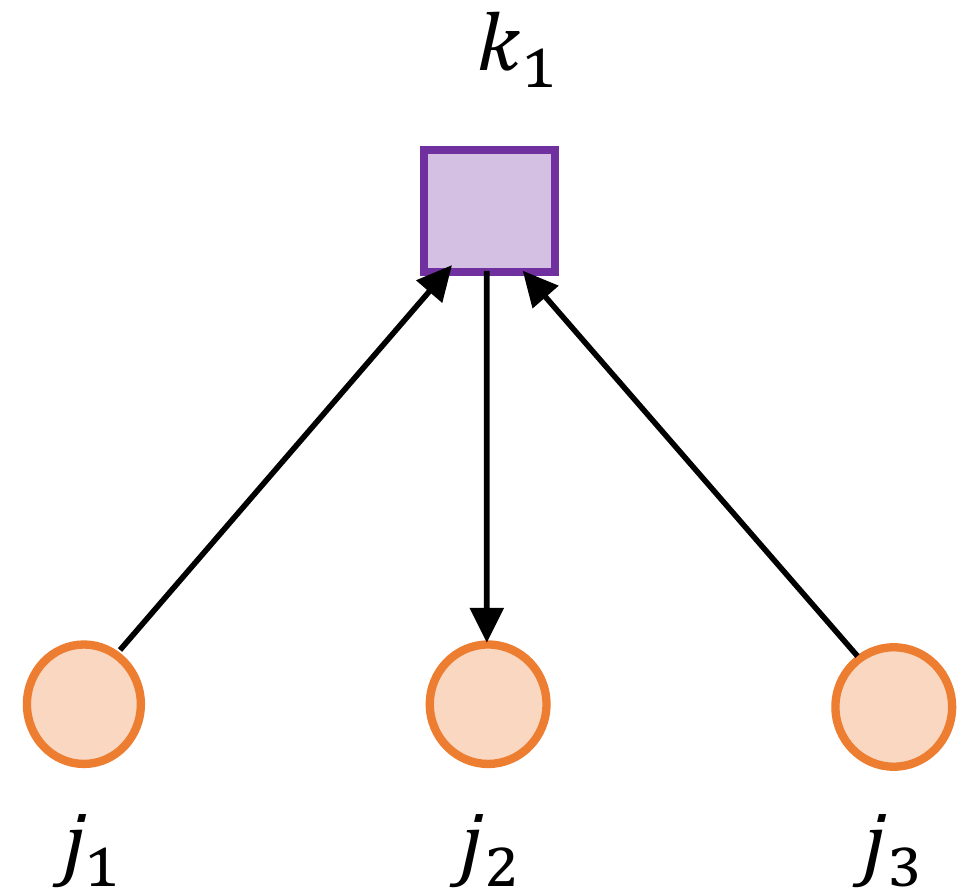}
\caption{Passing message from an FN to a VN. }
\label{fig:fn_vn}
\end{figure}

\end{itemize}
In this way, belief  messages are passed from every connected pair of FN and VN in both directions (FN $\Longleftrightarrow$ VN). Assuming that the  messages passed through the factor graph are statistically independent, exact APPs can be computed. In the case of cyclic factor graph (i.e., factor graph with cycles in it),  independent assumption is true for initial iterations only.  However in practice,  simulations have shown that MPA provides very effective results for computing APPs or LLRs even in cyclic factor graphs when the size of each cycle is large enough\cite{tarokh}.%at least 6. 
\subsection{Sum-Product Algorithm}
In the previous sub-section, we discussed how messages are passed from FN to VN and vice-versa. In this sub-section, we will explain with the help of an example how messages are passed across a factor graph and the resultant inference obtained. Consider a factor graph having $K$ function nodes (FNs) and $J$ variable nodes (VNs). The sum-product algorithm (SPA) starts by passing messages from leaf nodes. Let say a node has a degree (number of edges connected to it) $d$, then it will remain idle until  messages have arrived on $d-1$ edges.

Let $n_{j \rightarrow k}$  be the message sent from the  VN $j$ to the  FN $k$. Similarly, let $n_{k \rightarrow j}$ be the message from the FN $k$ to VN $j$. Let $\zeta_j$ and $R=\xi_k$ be the set of nodes directly connected to  VN $j$ and  FN $k$,  respectively. Let $A_j$ be the set of information symbols  that can be sent from VN $j$ and $a_j \in A_j$. Next, SPA can be carried out as follows: 
\begin{enumerate}
\item 	Passing message from FN $k$ to VN $j$:
\begin{equation}\label{fn_vn_spa}
\begin{split}
    n_{k \rightarrow j}(a_j) &=  \sum_{{\sim}a_j} \left(\psi(R) \prod_{r \in R \setminus \{j\}} n_{r \rightarrow k}(a_r)\right) \\
    &~~~~~~~~~~~~~~~~~~~~~~~~~~~~~~~\text{for} ~a_j \in A_j,
    \end{split}
\end{equation}
where $\psi(R)$ is the likelihood function associated with FN $k$ and $r \in R \setminus \{j\}$ denotes all the VNs of $R$ except VN $j$, respectively. Also, $n_{r \rightarrow k}(a_r)$ denotes the message from the VN $r$ to the FN $k$ corresponding to symbol $a_r$.
\item	Passing message from VN $j$ to FN $k$:
\begin{align}\label{vn_fn_spa}
    n_{j \rightarrow k}(a_j) =\prod_{d \in \zeta_j \setminus \{k\}} n_{d \rightarrow j}(a_j)~~~ \text{for} ~a_j \in A_j.
\end{align}
where $d \in \zeta_j \setminus \{k\}$ denotes the FNs in $\zeta_j$ except the $k$th FN. The message is normalized to ensure that the sum of all the probabilities is equal to 1.
\begin{align}\label{vn_fn_spaa}\nonumber
    n_{j \rightarrow k}(a_j) =\frac{\prod_{d \in \zeta_j \setminus \{k\}} n_{d \rightarrow j}(a_j)}{\sum_{a_j} \prod_{d \in \zeta_j \setminus \{k\}} n_{d \rightarrow j}(a_j)  } \\%\hspace{.5cm}
    \text{for} ~a_j \in A_j.
\end{align}

\end{enumerate}
The above algorithm makes use of the sum and product operations and hence it is called sum-product algorithm. The detailed explanation of (\ref{fn_vn_spa}) and (\ref{vn_fn_spa}) is given in \cite{facsum}. It is noted that  (\ref{vn_fn_spa}) is simpler than (\ref{fn_vn_spa}) because there is no local function associated with a VN. Next, we discuss how messages are passed over a factor graph using SPA and the resultant marginal function obtained with respect to each variable.

\begin{itemize}
    \item[$-$] \emph{Example 8:} 	Consider a global function $\phi(j_1,j_2,j_3,j_4)$ and the set of values  $j_i$ can take is $A_{j_i}$ and $a_{j_i} \in  {A}_{j_i}$. The global function can be written as a product of local functions (each having argument as subsets of $\{j_1,j_2,j_3,j_4$\}) as  
    \begin{align}\label{spa_ex}
        \phi(j_1,j_2,j_3,j_4 )= k_1 (j_1 ) k_2 (j_2,j_3 ) k_3 (j_1,j_3,j_4 ).
    \end{align}
%\end{itemize}

There is one marginal function associated with each variable, i.e. $\phi_i (a_{j_i})$ for $i=1, 2, 3, 4$. The  factor graph corresponding to   (\ref{spa_ex}) is shown in Fig. \ref{fig:spa_fac_graph}. It has 4 VNs corresponding to 4 variables  ( $j_1,j_2,j_3,j_4$) and 3 FNs corresponding to 3 local functions ($k_1,k_2,k_3$), respectively. An edge will be connected between a VN and a FN if that variable is an argument of that local function. For example, for $k_1(j_1)$, an edge will be present between FN $k_1$ and VN $j_1$. The passing of messages using SPA in the factor graph of Fig. \ref{fig:spa_fac_graph} is explained in the following steps:
      %   trim=left bottom right top
%\fbox{\includegraphics[scale=0.5,trim=6cm 6cm 9cm 6.5cm,clip]{spa_fac_gra.pdf}}
\begin{figure} 
\centering
\includegraphics[scale=0.4,trim=6cm 6cm 9cm 6.5cm,clip]{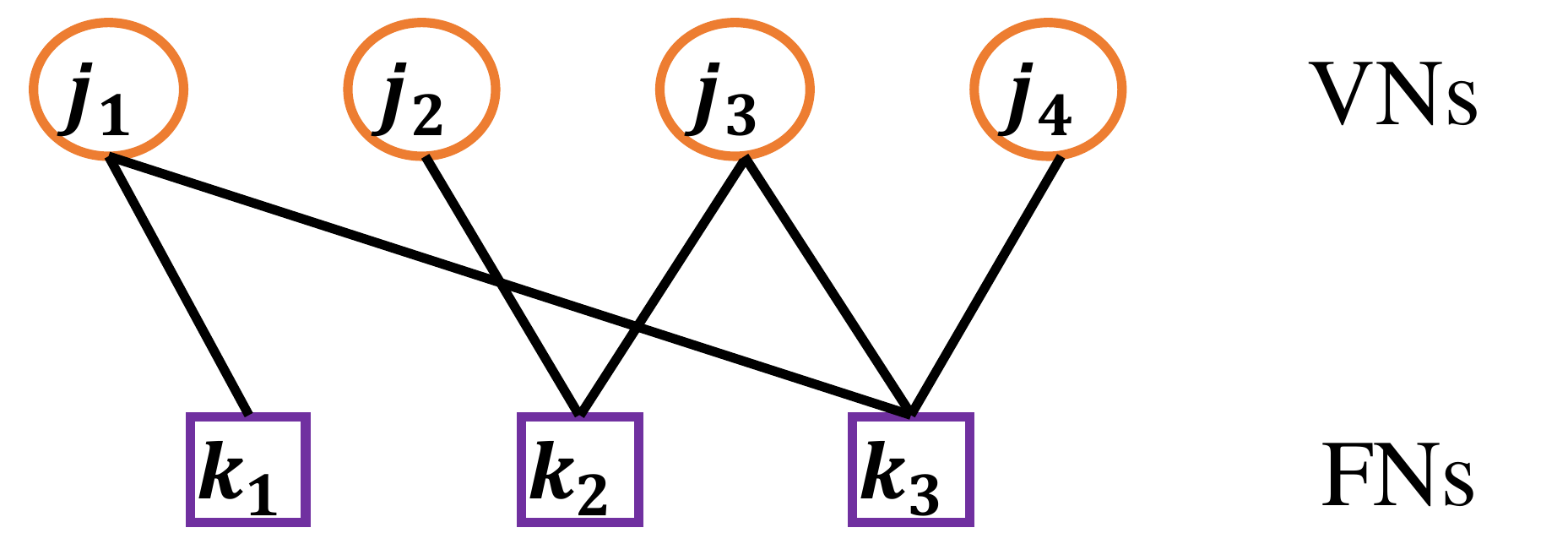}
\caption{Factor graph for the product $k_1 (j_1 ) k_2 (j_2,j_3 ) k_3 (j_1,j_3,j_4 ).$ }
\label{fig:spa_fac_graph}
\end{figure}

\begin{itemize}
\item[$\bullet$]{Step 1:}
SPA starts from the leaf nodes, i.e. $k_1,j_2 $ and $j_4$.
\begin{align*}
    & n_{k_1 \rightarrow j_1}(a_{j_1})  =k_1(a_{j_1}) \hspace{1.1cm} \forall~ a_{j_1} \in {A}_{j_1},\\
    & n_{j_2 \rightarrow k_2 } (a_{j_2} )= 1 \hspace{1.95cm} \forall ~a_{j_2} \in {A}_{j_2},\\
    &n_{j_4 \rightarrow k_3 } (a_{j_4} )=1 \hspace{2cm} \forall~ a_{j_4} \in {A}_{j_4}.
\end{align*}

The initial message from a VN to FN can be 1 as default.

\item[$\bullet$]{Step 2:} VN $j_1$ is connected to two FNs $k_1$ and $k_3$, i.e., VN $j_1$ has degree 2. Since VN $j_1$ has received message from FN $k_1$ in Step 1 (i.e. it has received message from $d-1$ edges), it can now pass message to FN $k_3$. 
\begin{equation*}
    \begin{split}
n_{j_1 \rightarrow k_3} (a_{j_1})=n_{k_1 \rightarrow j_1 } (a_{j_1}) \hspace{0.2cm} \forall~ a_{j_1} \in {A}_{j_1},\\
n_{k_2 \rightarrow j_3} (a_{j_3} )= \sum_{a_{j_2} \in {A}_{j_2}} (k_2(a_{j_2},a_{j_3}) ~ \\n_{j_2 \rightarrow k_2} (a_{j_2}))\hspace{0.2cm}
    \forall~ a_{j_3} \in {A}_{j_3}.
\end{split}
\end{equation*}
To pass message from FN $k_2$ to VN $j_3$, the message from VN $j_2$ to FN $k_2$ is multiplied with the local function of FN $k_2$ and then summation is performed with respect to all variables of FN $k_2$ except VN $j_3$. This results in an outgoing message in terms of variable $j_3$ only. When the belief message is passed from FN $k_2$ to VN $j_3$, the message  from VN $j_2$ to FN $k_2$ is called extrinsic information.
\item[$\bullet$]{Step 3:}
Once FN $k_3$ has received messages from VNs $j_1$ and $j_4$ (all neighbors except VN $j_3$), it will then generate and pass a message to VN $j_3$. Here, $n_{j_1 \rightarrow k_3} (a_{j_1} )$ and $n_{j_4 \rightarrow k_3} (a_{j_4} )$ are considered as extrinsic information.
\begin{equation*}
    \begin{split}
    &    n_{k_3 \rightarrow j_3} (a_{j_3})= \sum_{a_{j_1} \in {A}_{j_1}}  \sum_{a_{j_4} \in {A}_{j_4}}\ \biggl(k_3 (a_{j_1},a_{j_3},a_{j_4} )\\
&~~~~~\times  n_{j_1 \rightarrow k_3} (a_{j_1}) ~ n_{j_4 \rightarrow k_3} (a_{j_4})\biggr) ~~~ \forall~ a_{j_3} \in {A}_{j_3},\\ 
&    n_{j_3 \rightarrow k_3} (a_{j_3})= n_{k_2 \rightarrow j_3} (a_{j_3})\,~~~~~~~~~~\forall~ a_{j_3} \in {A}_{j_3}.
    \end{split}
\end{equation*}
     %   trim=left bottom right top
%\fbox{\includegraphics[scale=0.5,trim=4.75cm 2cm 3.75cm 7.75cm,clip]{step_spa.pdf}}
\begin{figure*}[htbp]
\centering
\includegraphics[scale=0.45,trim=4.75cm 2cm 3.75cm 7.75cm,clip]{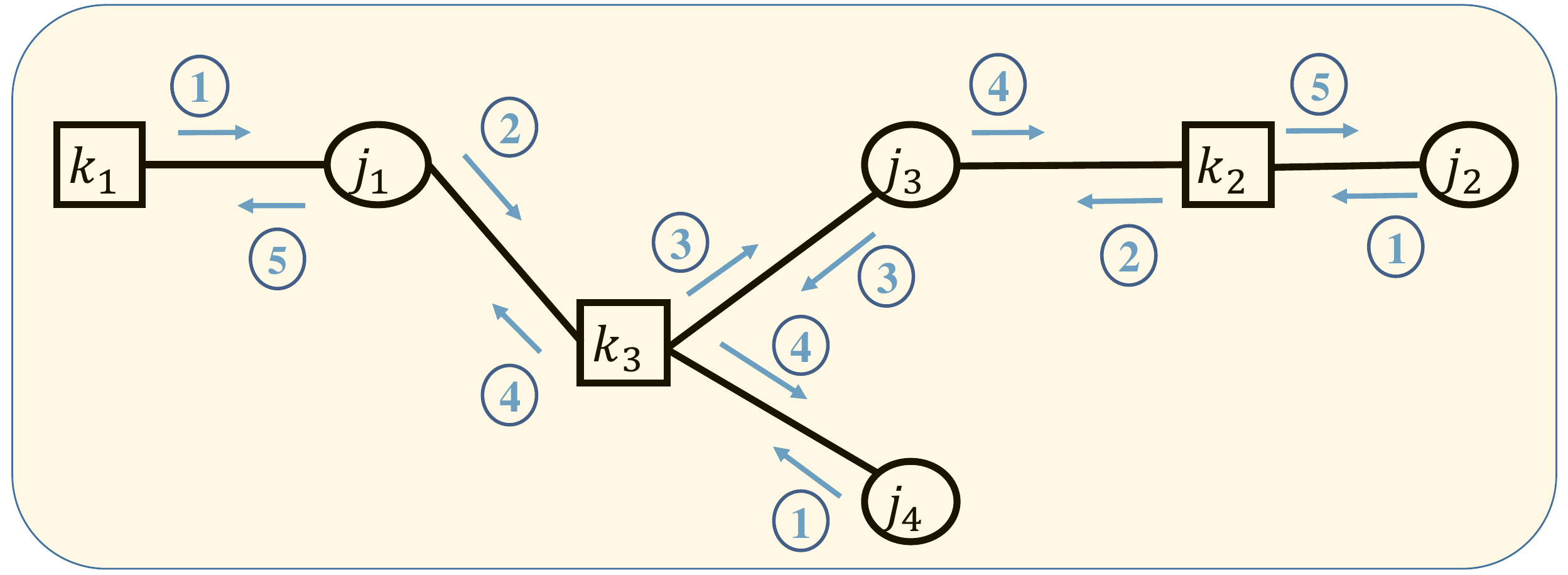}
\caption{Step-wise message passing in Factor graph.}
\label{fig:step_spa}
\end{figure*}
\item[$\bullet$]{Step 4:}
\begin{equation*}
\begin{split}
    & n_{k_3 \rightarrow j_1} (a_{j_1} )= \sum_{a_{j_3} \in {A}_{j_3}} 
    \sum_{a_{j_4} \in {A}_{j_4}} \biggl(k_3(a_{j_1},a_{j_3},a_{j_4} ) \\
    & ~~~~~ \times n_{j_3 \rightarrow k_3} (a_{j_3} ) ~ n_{j_4 \rightarrow k_3 } (a_{j_4} )\biggr)  ~~~\forall~ a_{j_1} \in {A}_{j_1},\\
&    n_{k_3 \rightarrow j_4} (a_{j_4})= \sum_{a_{j_1} \in {A}_{j_1}}  \sum_{a_{j_3} \in {A}_{j_3}}\ \biggl(k_3 (a_{j_1},a_{j_3},a_{j_4} )\\
&~~~~~ \times  n_{j_3 \rightarrow k_3} (a_{j_3}) ~ n_{j_1 \rightarrow k_3} (a_{j_1})\biggr) ~~~ \forall~ a_{j_4} \in {A}_{j_4},\\ 
&    n_{j_3 \rightarrow k_2} (a_{j_3})= n_{k_3 \rightarrow j_3} (a_{j_3})\,~~~~~~~~~~\forall~ a_{j_3} \in {A}_{j_3}.
\end{split}
\end{equation*}

\item[$\bullet$]{Step 5:}
\begin{equation*}
    \begin{split}
    &    n_{k_2 \rightarrow j_2} (a_{j_2})=  \sum_{a_{j_3} \in {A}_{j_3}}\ k_2 (a_{j_2},a_{j_3} ) n_{j_3 \rightarrow k_2} (a_{j_3})\\
&    ~~~~~~~~~~~~~~~~~~~~~~~~~~~~~~~~~~~~~~~~~~~~ \forall~ a_{j_2} \in {A}_{j_2},\\ 
&    n_{j_1 \rightarrow k_1} (a_{j_1})= n_{k_3 \rightarrow j_1} (a_{j_1})\,~~~~~~~~~~~\forall~ a_{j_1} \in {A}_{j_1}.
    \end{split}
\end{equation*}
The algorithm continues until messages are passed through all the edges in both directions. The order of messages generated when SPA is applied over the factor graph of Fig. \ref{fig:spa_fac_graph} is shown in Fig. \ref{fig:step_spa}. 
\end{itemize}
\emph{Termination}:
The marginal function of a VN can be  computed by multiplying all the received messages at the respective VN. For instance, the neighbouring nodes of VN $j_1$ are FNs $k_1$ and $k_3$, respectively. Therefore, the product of messages received from $k_1$ and $k_3$ will be the resultant message at the VN $j_1$. Similarly, the marginal function can be generated for other VNs.
\begin{align*}
   & \phi_1 (a_{j_1} )=n_{k_3 \rightarrow j_1} (a_{j_1}) ~ n_{k_1 \rightarrow j_1} (a_{j_1}) ~~~~ \hspace{0.05cm}\forall ~ a_{j_1} \in {A}_{j_1},\\
&    \phi_2 (a_{j_2} )=n_{k_2 \rightarrow j_2} (a_{j_2} ) ~~~~~~~~~~~~~~~~~~~~ \forall ~ a_{j_2} \in {A}_{j_2},\\
&    \phi_3 (a_{j_3})= n_{k_2 \rightarrow j_3} (a_{j_3} ) ~ n_{k_3 \rightarrow j_3} (a_{j_3} ) ~~~~ \hspace{0.05cm}\forall ~ a_{j_3} \in {A}_{j_3},\\
&    \phi_4 (a_{j_4})=n_{k_3 \rightarrow j_4} (a_{j_4}) ~~~~~~~~~~~~~~~~~~~~ \forall ~ a_{j_4} \in {A}_{j_4}.
\end{align*}
\end{itemize}

\subsection{SCMA Decoding}
In this section, with the help of the signature matrix of SCMA systems, we design the factor graph and use SPA to detect the symbols transmitted by each user.  In a factor graph,  each VN denotes an SCMA user and each FN denotes a  resource element. In the signature matrix  $\textbf{F}_{4\times6}$ given in (\ref{fac_gra_4_6}), there are  $d_v$ ones in each column and $d_f$ number of ones in each row, respectively.  In $\textbf{F}_{4\times6}$, every column corresponds to a user and every row corresponds to a RE, respectively. The first column of $\textbf{F}_{4\times6}$ indicates first user and it has non-zero values at first and second row. This means that data of first user is transmitted on the first and second RE, so there is an edge between VN $j_1$ and FN $k_1$ and, VN $j_1$ and FN $k_2$, respectively. The second row of $\textbf{F}_{4\times6}$ indicates second RE and it has non-zero values at first, fourth and fifth position. This means that data of first, fourth and fifth user overlaps on second RE and thus there will be edges connecting the FN $k_2$ with VN $j_1$, with VN $j_4$ and with VN $j_5$, respectively as shown in Fig. 12. %\ref{scma_fg}.

%   trim=left bottom right top
%\fbox{\includegraphics[scale=0.5,trim=2.75cm 6cm 5cm 6cm,clip]{scma_fac_gra.pdf}}

\begin{figure}[h!]
%\label{scma_fg}
\centering
\includegraphics[scale=0.32,trim=2.75cm 6cm 5cm 6cm,clip]{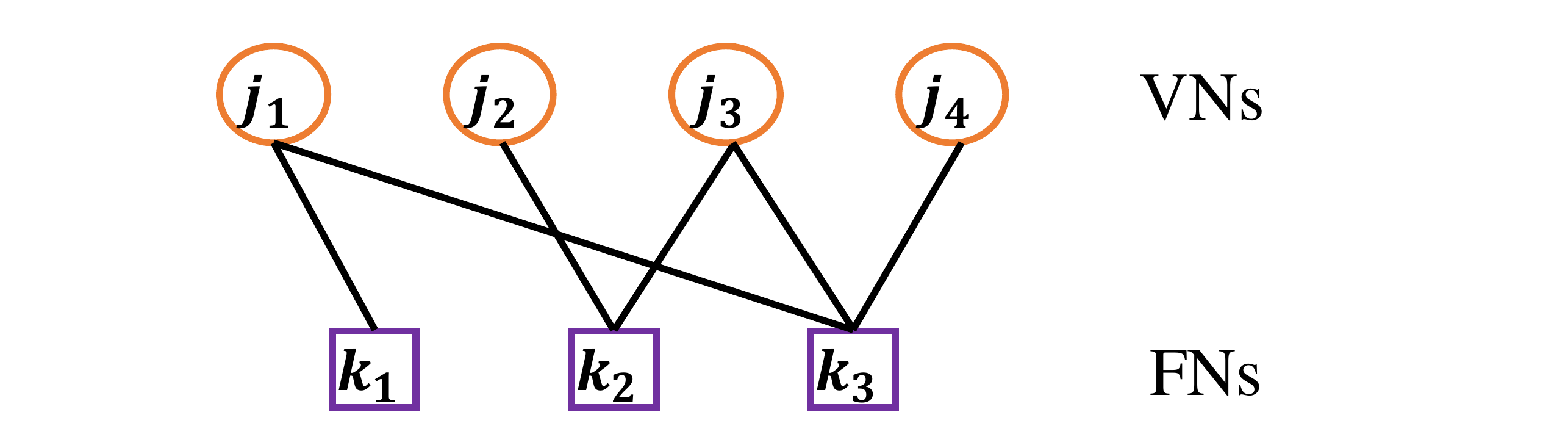}
\caption{Factor graph of the $4\times 6$ SCMA system corresponding to the signature matrix shown in (\ref{fac_gra_4_6}).}
\label{fig:scma_fac}
\end{figure}
Our aim is to detect the message transmitted by each user  using   (\ref{mpf}). To this end, SPA is applied over the factor graph shown in Fig. \ref{fig:scma_fac} and messages are passed between the VNs and FNs in both directions of the SCMA factor graph.  It is noted that  the factor graph of SCMA system contains cycles in it, and thus messages are passed between the nodes for a number of iterations until the termination criteria is achieved. 
\subsubsection{SPA for SCMA Decoding}

The objective is to detect the symbols transmitted by each user, i.e., compute (\ref{mpf}) using SPA and this  can be carried out in the following steps:
\begin{enumerate}
    \item Initialization.
    \item Passing of messages between FNs and VNs.
    \item Termination and selection of codewords.
\end{enumerate}
The algorithm is discussed in a detailed manner as follows:
\begin{itemize}
 \item Step 1: \textbf{Initialization}

 Assuming  received vector \textbf{y} and channel coefficient matrix \textbf{H} are known at receiver, firstly the likelihood ratio %$f(y_k \vert \textbf{X})$% 
 is computed at each FN. Let us assume a FN $l$ where data of three users corresponding to set $\xi_l=\{v_1.v_2,v_3\}$ superimpose. Therefore, the likelihood function at FN $l$ becomes $f(y_l \vert \textbf{m}_1,\textbf{m}_2,\textbf{m}_3,N_0)$, where, $\textbf{m}_1,\textbf{m}_2,\textbf{m}_3$ are the codewords transmitted by users belonging to set $\xi_l$, respectively. Assume the set of codewords allotted to user $v$ is denoted as $\mathbb{A}_v$. The likelihood function of FN $l$ is given as
\begin{align}\nonumber
        f(y_l \vert \textbf{m}_1,&\textbf{m}_2,\textbf{m}_3,N_0 )= \exp \biggl(\frac{-1}{N_0} ||y_l-(h_{l1}\\\nonumber C_{l,1} (\textbf{m}_1 )
       &+h_{l2} C_{l,2} (\textbf{m}_2 )+h_{l3} C_{l,3} (\textbf{m}_3 )) ||^2 \biggr) \, \\ 
    &\text{for}~ \textbf{m}_1 \in \mathbb{A}_1,\textbf{m}_2 \in \mathbb{A}_2,\textbf{m}_3 \in \mathbb{A}_3.
\end{align}
Here, $C_{l,v}(\textbf{m}_{v})$ denotes the codeword element transmitted by the $v$th  user when sending $\textbf{m}_{v}$ codeword  on the $l$th RE. In total, $KM^3$ values %\footnote{The likelihood ratio needs to be calculated for all combinations of symbols that user can send to neighboring RE.}
are stored for function $f(y_l \vert \textbf{m}_1,\textbf{m}_2,\textbf{m}_3,N_0 )$.  For an uncoded SCMA system, let us assume  equal prior probability for each codeword, i.e., $P(\textbf{m}_1 )=P(\textbf{m}_2 )=P(\textbf{m}_3 )= \frac{1}{M}$. Therefore, the  initial message passed from  VN $v_1,v_2,v_3$  to the $l${th} FN is 
 \begin{align}
     n_{v_1 \rightarrow l}^{\text{init}} (\textbf{m}_1 )=n_{v_2 \rightarrow l} ^{\text{init}} (\textbf{m}_2 )=n_{v_3 \rightarrow l}^{\text{init}} (\textbf{m}_3 )=\frac{1}{M}\,.
 \end{align}

\item Step 2: \textbf{Passing of messages between FNs and VNs.}

%\begin{itemize}
    
     	\textbf{(i) From Function Node to Variable Node}:\\
    
    Let us assume  $\xi_l=\{v_1,v_2,v_3\}$, where $v_1,v_2,v_3$ denotes the three users connected to RE $l$, respectively.  To pass the message  from the FN to one user, information received on the FN from the other two users may be regarded as extrinsic information. 
    %   trim=left bottom right top
%\fbox{\includegraphics[scale=0.5,trim=3.75cm 6cm 4.5cm 6cm,clip]{fn_to_vn_scma.pdf}}

\begin{figure}
\centering
\includegraphics[scale=0.3,trim=3.75cm 6cm 4.5cm 6cm,clip]{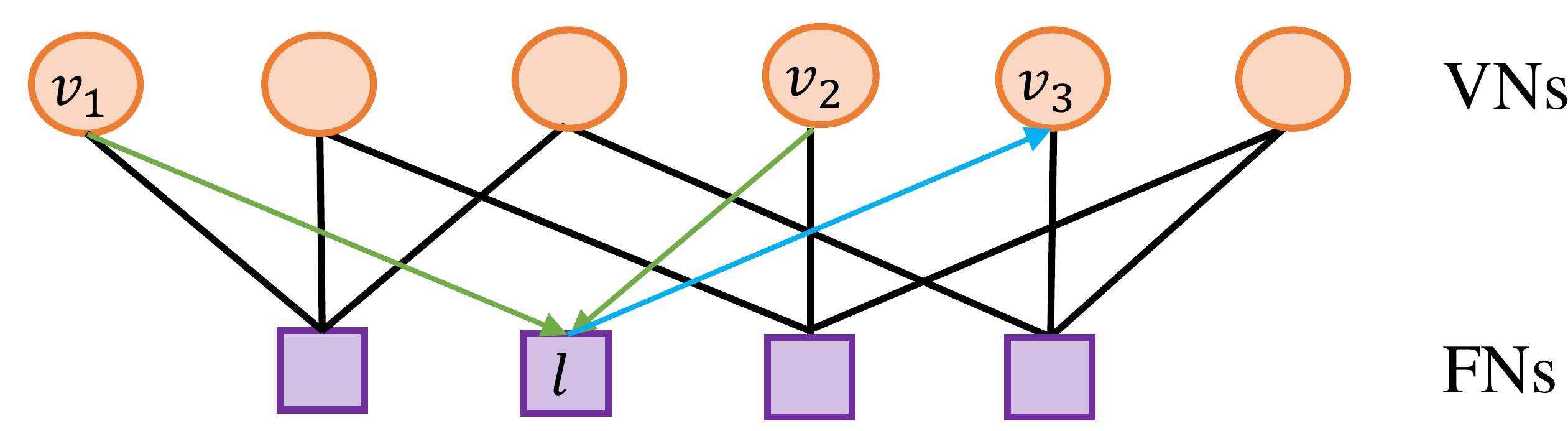}
\caption{Message Passing from FN to VN.}
\label{fig:scma_fn_vn}
\end{figure}
    
Message passed from FN $l$ to VN $v_1$ is given as
    \begin{multline}\label{Fn_vn_spa}
   % \begin{split}
        n_{l \rightarrow v_1} (\textbf{m}_1)\\= \sum_{\textbf{m}_2\in \mathbb{A}_2} \sum_{\textbf{m}_3\in \mathbb{A}_3} \biggl(f(y_l \vert \textbf{m}_1,\textbf{m}_2,\textbf{m}_3,N_0 )\\ 
\times         n_{v_2 \rightarrow l} (\textbf{m}_2 ) \hspace{0.05cm} n_{v_3 \rightarrow l} (\textbf{m}_3)\biggr) \hspace{0.25cm} \text{for} ~ \textbf{m}_1 \in \mathbb{A}_1.
      %  \end{split}
    \end{multline}
In equation (\ref{Fn_vn_spa}), messages from the two VNs ($n_{v_2 \rightarrow l} (\textbf{m}_2 )$ and $n_{v_3 \rightarrow l} (\textbf{m}_3 )$) are multiplied with the local likelihood  function of $l$th FN and then marginalized with respect to $v_1$. Similarly, message passed from FN $l$ to VNs $v_2$ and $v_3$, respectively are 

\begin{multline}
    %\begin{split}
    n_{l \rightarrow v_2} (\textbf{m}_2 )\\= \sum_{\textbf{m}_1\in \mathbb{A}_1} \sum_{\textbf{m}_3\in \mathbb{A}_3} \biggl(f(y_l \vert \textbf{m}_1,\textbf{m}_2,\textbf{m}_3,N_0 )\\ 
\times      n_{v_1 \rightarrow l} (\textbf{m}_1 ) \hspace{0.05cm} n_{v_3 \rightarrow l} (\textbf{m}_3)\biggr) ~~ \text{for} ~ \textbf{m}_2 \in \mathbb{A}_2. 
   % \end{split}
\end{multline}

\begin{multline}
   % \begin{split}
    n_{l \rightarrow v_3} (\textbf{m}_3 )\\= \sum_{\textbf{m}_1\in \mathbb{A}_1} \sum_{\textbf{m}_2\in \mathbb{A}_2} \biggl(f(y_l \vert \textbf{m}_1,\textbf{m}_2,\textbf{m}_3,N_0 )\\ 
\times     n_{v_1 \rightarrow l} (\textbf{m}_1 )\hspace{0.05cm}  n_{v_2 \rightarrow l} (\textbf{m}_2)\biggr)\, ~~ \text{for} ~\textbf{m}_3 \in \mathbb{A}_3.
%\end{split}
\end{multline}
Message from FN $l$ to VN $v_3$ is shown graphically in Fig. \ref{fig:scma_fn_vn}. The message transmitted from FN $l$ to VN $v$ indicates the guess of what signal is received at FN $l$ for all possible values of VN $v$.\\

  \textbf{ (ii)From Variable Node to  Function Node}:\\
 Let us assume $\zeta_v=\{l_1,l_2\}$, where $l_1$ and $l_2$ are the FNs connected to VN $v$.\\
%   trim=left bottom right top
%\fbox{\includegraphics[scale=0.5,trim=3.75cm 6cm 4.5cm 6cm,clip]{vn_to_fn_scma.pdf}}

 \begin{figure}[h!]
  \centering
\includegraphics[scale=0.3,trim=3.75cm 6cm 4.5cm 6cm,clip]{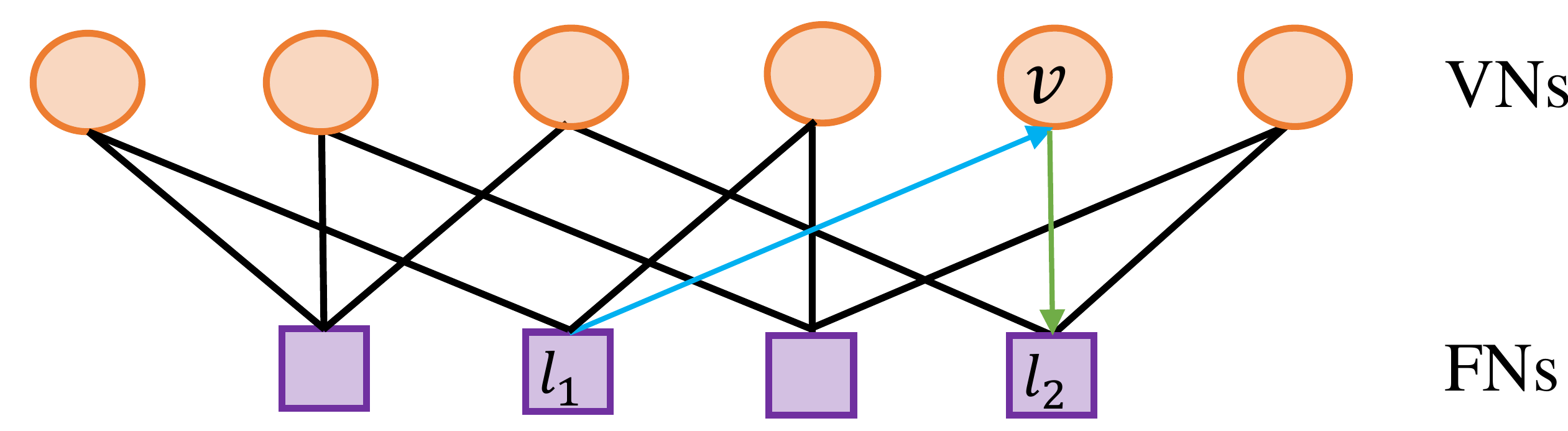}
\caption{Message Passing from VN to FN.}
\label{fig:scma_vn_fn}
\end{figure}
Message sent from VN $v$ to FN $l_1$ and $l_2$, respectively are

\begin{multline}\label{vn_fn_1}
    n_{v \rightarrow l_1}(\textbf{m}_v)= \text{normalize}(p_a (\textbf{m}_v) \hspace{0.1cm} n_{l_2 \rightarrow v }(\textbf{m}_v)) \\
    \text{for}~ \textbf{m}_v \in \mathbb{A}_v ,
\end{multline}

\begin{multline}\label{vn_fn_2}
            n_{v \rightarrow l_2}(\textbf{m}_v)= \text{normalize}(p_a (\textbf{m}_v) \hspace{0.1cm} n_{l_1 \rightarrow v} (\textbf{m}_v))\, \\
            \hspace{1cm} \text{for}~ \textbf{m}_v \in \mathbb{A}_v.
\end{multline}
where $p_a$ denotes the  prior probability of user $v$ and $n_{v \rightarrow l_2}$ indicates the updates VN $v$ received from the other FNs connected to $v$.
Here, normalization is necessary in order to ensure  that each belief falls in the range of [0,1].
The normalization is conducted similar to  (\ref{vn_fn_spaa}) and so,  (\ref{vn_fn_1}) can also be written as
\begin{align}\nonumber
    n_{v \rightarrow l_1}(\textbf{m}_v) =\frac{ p_a(\textbf{m}_v) n_{l_2 \rightarrow v}(\textbf{m}_v)}{\sum_{\textbf{m}_v}  n_{l_2 \rightarrow v}(\textbf{m}_v)  } \\ \text{for} ~\textbf{m}_v \in \mathbb{A}_v.
\end{align}
The message from the VN $v$ to FN $l_2$ is shown graphically in Fig. \ref{fig:scma_vn_fn}.
%\end{itemize}
Since factor graph has cycles in it, Step 2 is repeated for each iteration. This continues until no considerable change is observed in the belief computed at each VN. The number of iterations for which step 2 is repeated is denoted as $N_{\text{iter}}$.

 \item   Step 3:\textbf{ Termination and selection of codewords.}

After repeating Step 2 for  $N_{\text{iter}}$ number of iterations, the final belief is  computed at each VN which is the product of the prior probability and messages from the neighboring FNs of each VN.
\begin{multline}    \label{term}
    	I_v(\textbf{m}_v)= p_a(\textbf{m}_v) \hspace{0.05cm}  n_{l_1 \rightarrow v} (\textbf{m}_v) \hspace{0.05cm} n_{l_2 \rightarrow v} (\textbf{m}_v)\,  \\
    	\text{for}~ \textbf{m}_v \in \mathbb{A}_v.
\end{multline}
The probability for each codeword is calculated at each VN and the codeword  with highest probability becomes the estimated codeword for that user. This is one method of estimating the transmitted symbol for each user. Another method is to compute bit LLR from $I_v(\textbf{m}_v)$. Let there be $b$ bits per symbol, then the $b_i$th bit LLR for VN $v$ is
\begin{equation} \label{bit_llr}
\begin{split}
 LLR_{b_i}^v & =\log \frac{P(b_i=0)}{P (b_i=1)}\,, \\
                    & =\log \frac{\sum_{\textbf{m}_v \vert b_{i}=0} I_v(\textbf{m}_v)} {\sum_{\textbf{m}_v \vert b_{i}=1} I_v(\textbf{m}_v)}\,.
\end{split}
\end{equation}
%   trim=left bottom right top
%\fbox{\includegraphics[scale=0.5,trim=10cm 4cm 12cm 7cm,clip]{llr_bit.pdf}}
\begin{figure}
\centering
\includegraphics[scale=0.5,trim=10cm 4cm 12cm 7cm,clip]{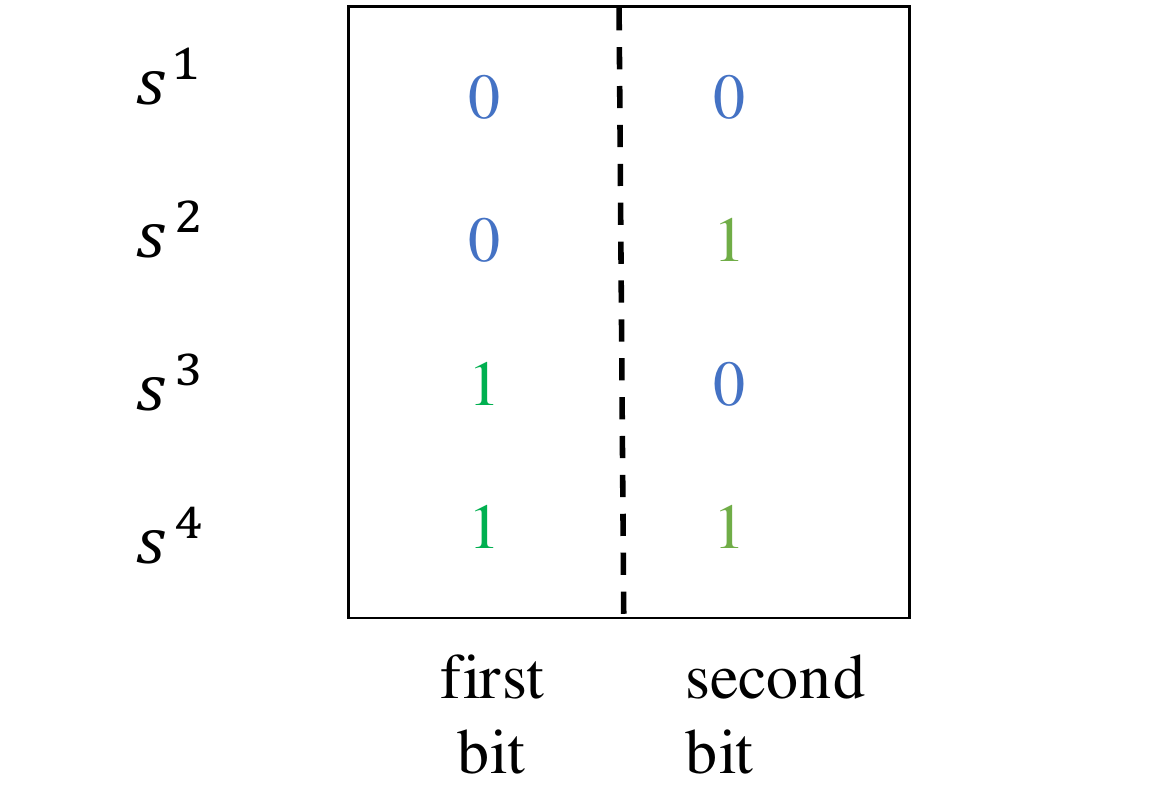}
\caption{Bit by bit decoding.}
\label{fig:llr_bitbit}
\end{figure}
In case of $M=4$, there are two bits per symbol. Let $\mathbb{A}_v=\{\textbf{m}_v^1,\textbf{m}_v^2,\textbf{m}_v^3,\textbf{m}_v^4\}$ denote the 4 codewords corresponding to 4 symbols $\{s^1,s^2,s^3,s^4\}$ that can be sent by user $v$. Using  (\ref{term}), final belief corresponding to each of the four codewords is calculated. %It is to be noted that 4 codewords, i.e. $\textbf{m}_1,\textbf{m}_2,\textbf{m}_3,\textbf{m}_4$ corresponds to the 4 symbols $s_1,s_2,s_3,s_4$ that can be sent by user $v$. 
Then,  the LLR of the first bit is given as
\begin{equation} 
\begin{split}
 LLR_{b_1}^v &=\log ⁡\frac{P (b_1=0)}{P (b_1=1)}\,, 
                       =\log ⁡\frac{ I_v(\textbf{m}_v^1 )+I_v(\textbf{m}_v^2)}{I_v(\textbf{m}_v^3)+I_v(\textbf{m}_v^4)}\,.
\end{split}
\end{equation}
It can be seen from Fig. \ref{fig:llr_bitbit} that first bit ($b_1$) is 0 for symbols $s^1$ and $s^2$ and is 1 for symbols $s^3$ and $s^4$, respectively. As such, the ratio of the belief computed corresponding to codewords $\textbf{m}_v^1,\textbf{m}_v^2$ and $\textbf{m}_v^3, \textbf{m}_v^4$ gives the bit LLR of the first bit. Consequently if $P(b_1=1)>P(b_1=0)$ or if $LLR_{b_1}^v<0$, then `1' is decoded for $1${st} bit of the symbol otherwise 0. Similarly for second bit, if $LLR_{b_2}^v<0$, then `1' is decoded for the second bit of the symbol.
\begin{align}
    LLR_{b_2}^v= \log ⁡\frac{ I_v(\textbf{m}_v^1 )+I_v(\textbf{m}_v^3)}{I_v(\textbf{m}_v^2)+I_v(\textbf{m}_v^4)}\,.
\end{align}
\end{itemize}

\subsubsection{Max-Log-SPA for SCMA Decoding}
The SPA algorithm discussed above computes LLR in probability domain. Although, it is simpler than MAP algorithm, it still has complexity issues mainly because of the message passing from FN to VN ($O (N_{\text{iter}} K M^{d_f})$). It involves number of product and exponential operations rendering it computationally intensive. Also, exponential operations involve large dynamic range which puts a burden on storage. Thus, similar to the max-log-MAP algorithm \cite{mapbook}, max-log-MPA algorithm \cite{mpa_max} can be used to further simplify the MPA. 

To this end, the SPA algorithm is moved into the logarithm domain and Jacobian logarithm  can also be used to avoid exponential operations, which further simplifies the algorithm, i.e.,
\begin{align}\label{jacob}
    \log (\exp (a_1)+\cdots+\exp(a_J)) \approx \text{max}(a_1,\cdots,a_J).
\end{align}
The max-log-SPA algorithm for (4,6) SCMA block is illustrated in the following steps:
\begin{itemize}
    
\item Step 1: \textbf{ Initialization.}\\
    Given $\textbf{y}$ and $\textbf{H}$, initial value of LLR is calculated.
    \begin{multline}
     \log (f(y_l \vert \textbf{m}_1,\textbf{m}_2,\textbf{m}_3,N_0 ))=  \frac{-1}{N_0} ||y_l-(h_{l1}\\ C_{l1} (\textbf{m}_1 )
     +h_{l2} C_{l2} (\textbf{m}_2 )+h_{l3} C_{l3} (\textbf{m}_3 )) ||^2 \, \\ \text{for}~ \textbf{m}_1 \in \mathbb{A}_1,\textbf{m}_2 \in \mathbb{A}_2,\textbf{m}_3 \in \mathbb{A}_3.    
    \end{multline}
The initial message from VN $v$ to FN $l$ is
    \begin{align}
        \eta_{v \rightarrow l}^{\text{init}} = log (n_{v \rightarrow l}^{\text{init}}) = \log (\frac{1}{M})\,.
    \end{align}
\item Step 2:\textbf{ Passing of Messages between FNs and VNs.}
  %  \begin{itemize}
     \textbf{(i) From Function Node to Variable Node.}\\
    Applying natural logarithm to  (\ref{Fn_vn_spa}). Let $a_i$ be
    \begin{align} \nonumber
   % \begin{multline}
    \begin{split}
        a_i = \log(f(y_l \vert \textbf{m}_1,\textbf{m}_2,\textbf{m}_3,N_0 ) &\hspace{0.05cm} n_{v_2 \rightarrow l} (\textbf{m}_2 ) \hspace{0.05cm} \\&n_{v_3 \rightarrow l} (\textbf{m}_3))\, \\
         = \log(f(y_l \vert \textbf{m}_1,\textbf{m}_2,\textbf{m}_3,N_0 ))+& \hspace{0.05cm} \eta_{v_2 \rightarrow l} (\textbf{m}_2 ) \hspace{0.05cm} + \\&
        \eta_{v_3 \rightarrow l} (\textbf{m}_3)\,.
    \end{split}
   % \end{multline}
    \end{align}
    
Using  $a_i$ as given in  (\ref{jacob}), now message passed from the $l$th FN to VN $v_1$ is given as:
\begin{multline}
    \eta_{l \rightarrow v_1}^{\text{init}}(\textbf{m}_1) \\= \max_{\textbf{m}_2,\textbf{m}_3} \biggl(\log(f(y_l \vert \textbf{m}_1,\textbf{m}_2,\textbf{m}_3,N_0 ))\\ + \eta_{v_2 \rightarrow l} (\textbf{m}_2 ) \hspace{0.05cm} + \eta_{v_3 \rightarrow l} (\textbf{m}_3)\biggr)\, \hspace{0.2cm}  \text{for}~ \textbf{m}_1 \in \mathbb{A}_1.
\end{multline}

Similarly, message from FN $l$ to VN $v_2$ and $v_3$, respectively are
\begin{multline}
    \eta_{l \rightarrow v_2}^{\text{init}}(\textbf{m}_2) \\= \max_{\textbf{m}_1,\textbf{m}_3} \biggl(\log(f(y_l \vert \textbf{m}_1,\textbf{m}_2,\textbf{m}_3,N_0 ))\\ + \eta_{v_1 \rightarrow l} (\textbf{m}_1 ) \hspace{0.05cm} + \eta_{v_3 \rightarrow l} (\textbf{m}_3)\biggr)\, \hspace{0.2cm} \text{for}~ \textbf{m}_2 \in \mathbb{A}_2.
\end{multline}
    
\begin{multline}
    \eta_{l \rightarrow v_3}^{\text{init}}(\textbf{m}_3) \\= \max_{\textbf{m}_1,\textbf{m}_3}  \biggl(\log(f(y_l \vert \textbf{m}_1,\textbf{m}_2,\textbf{m}_3,N_0 ))\\ + \eta_{v_1 \rightarrow l} (\textbf{m}_1 ) \hspace{0.05cm} + \eta_{v_2 \rightarrow l} (\textbf{m}_2)\biggr)\, \hspace{0.2cm} \text{for}~ \textbf{m}_3 \in \mathbb{A}_3.
\end{multline}

\textbf{(ii) From Variable Node to  Function Node.}\\
Since each user is sending data  on two REs, message passed from the $v$th VN to FNs $l_1$ and $l_2$ in logarithm domain, respectively are (applying log to  (\ref{vn_fn_1}-\ref{vn_fn_2}))
\begin{align}\nonumber
    \eta_{v \rightarrow l_1} (\textbf{m}_v) = \log (\frac{1}{M}) + \eta_{l_2 \rightarrow v}(\textbf{m}_v) \\ \text{for}~ \textbf{m}_v \in \mathbb{A}_v.
\end{align}
\begin{align}\nonumber
    \eta_{v \rightarrow l_2} (\textbf{m}_v) = \log (\frac{1}{M}) + \eta_{l_1 \rightarrow v}(\textbf{m}_v) \hspace{0.5cm}  \\\text{for}~ \textbf{m}_v \in \mathbb{A}_v.
\end{align}
Here, for simplicity normalization operation can be ignored, since exponential operations are no longer present.

%    \end{itemize}
\item Step 3: \textbf{Termination and selection of codewords.}\\    
    Step 2 is repeated for $N_{\text{iter}}$ iterations, then APP in the log domain is calculated for each codeword of each user. Applying logarithm to  (\ref{term}), we obtain
\begin{multline}
        	\log(I_v(\textbf{m}_v))= \log(\frac{1}{M}) \hspace{0.05cm} + \eta_{l_1 \rightarrow v} (\textbf{m}_v) \hspace{0.05cm} \\+ \eta_{l_2 \rightarrow v} (\textbf{m}_v)  
        	~~~~~ \text{for}~ \textbf{m}_v \in \mathbb{A}_v.	
\end{multline}
    
Using  (\ref{bit_llr}) and (\ref{jacob}), bit LLR of $b_i${th} bit is given as
\begin{align}\nonumber
    LLR_{b_i}^v= \max_{\textbf{m}_v \vert b_i=0} (\log(I_v(\textbf{m}_v)))  \\- \max_{\textbf{m}_v \vert b_i=1} (\log(I_v(\textbf{m}_v))).
\end{align}    
The LLR of first bit of the symbol is
    \begin{multline}
    LLR_{b_1}^v= \max_{\textbf{m}_v^1,\textbf{m}_v^2} (\log(I_v(\textbf{m}_v^1 )+I_v(\textbf{m}_v^2 )) \enspace\\- \max_{\textbf{m}_v^3,\textbf{m}_v^4} (\log(I_v(\textbf{m}_v^3)+I_v(\textbf{m}_v^4))).
\end{multline}

    If $LLR_{b_1}^v$ $<$ 0, then `1' is decoded for the first bit of the symbol otherwise 0. Similarly, bit LLR for the second bit is given as
    \begin{multline}
    LLR_{b_2}^v= \max_{\textbf{m}_v^1,\textbf{m}_v^3} (\log(I_v(\textbf{m}_v^1 )+I_v(\textbf{m}_v^3 )) \enspace \\
    ~~ - \max_{\textbf{m}_v^2,\textbf{m}_v^4} (\log(I_v(\textbf{m}_v^2)+I_v(\textbf{m}_v^4))).
    \end{multline}
Again if $LLR_{b_2}^v$ is less than 0, `1' is decoded for the second bit of the symbol otherwise zero.
\end{itemize}    
We can see that in max-log-MPA,  exponential operations are no longer needed and more than $90\% $ of multiplication operations are removed with respect to SPA. Although, it involves many addition operations but overall it has much lower complexity than SPA. 
\begin{comment}
\begin{figure}
\centering
\includegraphics[scale=0.7]{mpa,max-log-mpa comparison,M=4.png}
\caption{Performance Comparison of SPA and Max-Log-SPA for (4,6) SCMA Block in Rayleigh Fading Channel for 4 QAM}
\label{fig:M_4}
\end{figure}    
\end{comment}

\begin{comment}
\begin{figure}
\centering
\includegraphics[scale=0.7]{mpa,max-log-mpa comparison,M=16.png}
\caption{Performance Comparison of SPA and Max-Log-SPA for (4,6) SCMA Block in Rayleigh Fading Channel for 16 QAM}
\label{fig:M_16}
\end{figure}  
\end{comment}
%   trim=left bottom right top
%\fbox{\includegraphics[scale=0.75,trim=3cm 8.5cm 3cm 9cm,clip]{scma_mpa_max_logmpa_oma_qpsk_div_ord1_ord2_ebno_30.pdf}}

\begin{figure}[h!]
\centering
\includegraphics[scale=0.55,trim=3cm 8.5cm 3cm 9cm,clip]{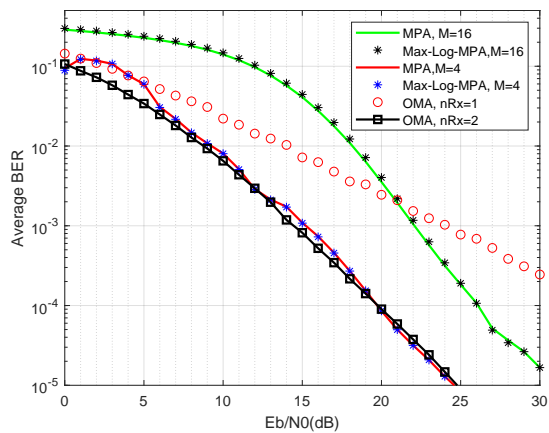}
\caption{Performance Comparison of SPA and Max-Log-SPA for (4,6) SCMA Block in Rayleigh Fading Channel for M=4, M=16 and OMA for Diversity order 1 and 2.}
\label{fig:M_4_16}
\end{figure}  

    Fig. \ref{fig:M_4_16} shows average bit error rate (BER) obtained of SCMA for M=4 and M=16 in Rayleigh Channel using SPA and Max-Log-SPA algorithms. It can be observed that there is negligible difference in the performance of the two algorithms. In Fig. \ref{fig:M_4_16}, we compare it with the BER performance of OMA scheme with diversity order 1 and 2 which has no multi-user interference. It can be noted that in SCMA data of six users is being transmitted using four subcarriers and suffers multi-user interference, still its performance is comparable with  OMA scheme of diversity order 2. Also, it was found in simulations that the run time in Max-Log-SPA is $~20\%$ and $~30\%$ lesser than SPA for M=4 and M=16 respectively. Therefore, practically Max-Log-SPA is a preferable solution due to its low complexity and near-optimal performance.
    
%\end{itemize}
 
\section{Conclusions}
%This paper presents a study on sum-product algorithm and the SCMA system model. Motivated by the fact that 5G and beyond wireless networks will support a diverse range of traffic, % with requirements like massive connectivity, spectral efficiency and low latency, 
In this paper, we have provided a systematic self-contained tutorial to SCMA which is a disruptive code-domain NOMA scheme for the enabling of massive connectivity.  
SCMA uses carefully designed sparse codebooks for significant constellation shaping gain and anti-interference capability as compared to previously proposed code-domain NOMA techniques. In comparison to the complex MAP decoding, iterative message passing algorithm gives rise to efficient MPA decoding with significantly reduced complexity. Thus, SCMA represents a promising solution for  providing better quality of service to users (with overloading factor greater than 1), low latency and high spectral efficiency. It is anticipated that this paper will provide a quick and comprehensive understanding of SCMA to motivate more researchers towards this research topic.

%In this paper, we discussed the  SCMA  system  using the message passing algorithm. The concepts required for the understanding of SCMA, i.e. soft decoding, factor graph and sum-product algorithm have been discussed with examples. Two types of decoding algorithms have been discussed, namely SPA and max-log-SPA  and it was noted that the latter achieves significant complexity reduction with negligible performance loss.
%However, it faces challenges in hardware implementation when the overloading factor is large.
%The concept and decoding mechanism of SCMA have been explained in a simplified manner to ease the  understanding of the topic. It is anticipated that this paper will provide a quick and comprehensive understanding of the SCMA system model motivating more researchers towards this research topic. 

\bibliographystyle{IEEEtran}
%\bibliography{references}

\end{document}